\documentstyle[12pt]{article}
\newcommand{ \R} {\mbox{\rm I$\!$R}}
\begin{document}

\title{Locally Anisotropic Kinetic Processes and \\
Thermodynamics in Curved Spaces}
\author{Sergiu I.\ Vacaru \thanks{
e--mails: svacaru@phys.asm.md, sergiu$_{-}$vacaru@yahoo.com, 
 vacaru@lises.asm.md } \quad \\
%EndAName
{\small Fachbereich Physik, Universitat Konstanz, }\\
{\small Postfach M 638, D--78457, Konstanz, Germany }\\
{\small and }\\
{\small Institute of Applied Physics, Academy of Sciences, }\\
{\small 5 Academy str., Chi\c sin\v au MD2028, Republic of Moldova}}
\date{August 13, 2000}
\maketitle

\begin{abstract}
The kinetic theory is formulated with respect to anholonomic frames of
reference on curved spacetimes. By using the concept of nonlinear connection
we develop an approach to modelling locally anisotropic kinetic processes
and, in corresponding limits, the relativistic non--equilibrium
thermodynamics with local anisotropy. This lead to a unified formulation of
the kinetic equations on (pseudo) Riemannian spaces and in various higher
dimensional models of Kaluza--Klein type and/or generalized Lagrange and
Finsler spaces. The transition rate considered for the locally anisotropic
transport equations is related to the differential cross section and
spacetime parameters of anisotropy. The equations of states for pressure and
energy in locally anisotropic thermodynamics are derived. The obtained
general expressions for heat conductivity, shear and volume viscosity
coefficients are applied to determine the transport coefficients of cosmic
fluids in spacetimes with generic local anisotropy. We emphasize that such
locally anisotropic structures are induced also in general relativity if we
are modelling physical processes with respect to frames with mixed sets of
holonomic and anholonomic basis vectors which naturally admits an associated
nonlinear connection structure.
\end{abstract}

\vskip0.3cm gr-qc/0001060;\ accepted to {\it Annals of Physics (NY)} \vskip%
0.5cm %\tableofcontents
%%%%%%%%%%%%%%%%%%%%%%%%%%%%%%%%%%%%%%%%%%%%%%%%%%%%%%%%%%%%%%%%%%%%
\newpage

\section{Introduction}

The experimental data on anisotropies in the microwave background radiation
(see, for instance, Ref. \cite{smoot} and \cite{ganga}) and modern physical
theories support the idea that in the very beginning the Universe has to be
described as an anisotropic and higher dimension spacetime. It is of
interest the investigation of higher dimension generalized Kaluza--Klein and
string anisotropic cosmologies. Therefore, in order to understand the
initial dynamical behavior of an anisotropic universe, in particular, to
study possible mechanisms of anisotropic inflation connected with higher
dimensions \cite{v3} we have to know how we can compute the parameters
(transport coefficients, damping terms and viscosity coefficients) that
characterize the cosmological fluids in spacetimes with generic anisotropy.

The first relativistic macroscopic thermodynamic theories have been proposed
in Refs. \cite{eckart} and \cite{landau}. The further developments and
applications in gravitational physics, astrophysics and cosmology are
connected with papers \cite{thomas,weinberg,hawking,neugebauer}. The
Israel's approach to a microscopic Boltzmann like kinetic theory for
relativistic gases \cite{israel} makes possible to express the transport
coefficients via the differential cross sections of the fluid's particles.
Here we also note the Chernikov's results on Boltzmann equations with
collision integral on (pseudo) Riemannian spaces \cite{chernikov}. A
complete formulation of relativistic kinetics is contained in the monograph
\cite{groot2} by de Groot, van Leeuwen and van Weert. We also emphasize the
Vlasov's monograph \cite{vlas} were an attempt to statistical motivation of
kinetic and thermodynamic theory on phase spaces enabled with Finsler like
metrics and connection structures was proposed. Recent developments and
applications of kinetics and nonequilibrium thermodynamics could be found in
\cite{zimd}.

The extension of the four dimensional considerations to higher dimensions is
due to \cite{tomita} and \cite{boisseau}. The generalization of kinetic and
thermodynamic equations and formulas to curved spaces is not a trivial task.
In order to consider flows of particles with noninteger spins, interactions
with gauge fields, and various anisotropic processes of microscopic or
macroscopic nature it is necessary a reformulation of the kinetic theory in
curved spacetimes by using the Cartan's moving frame method \cite{cartan1}.
A general approach with respect to anholonomic frames contains the
possibility to take into account generic spacetime anisotropies which play
an important role in the vicinity of cosmological or astrophysical
singularities and for non--trivial reductions of higher dimension theories
to lower dimensional ones. In our works \cite{v1,v2,v3} we proposed to
describe such locally anisotropic spacetimes and interactions by applying
the concept of nonlinear connection (in brief, N--connection) \cite{barthel}
field which models the local splitting of spacetime into horizontal
(isotropic) and vertical (anisotropic) subspaces. For some values of
components the N--connection could parametrize, for instance, toroidal
compactifications of higher dimensions but, in general, its dynamics is to
be determined, in different approaches, by field equations or constraints in
some generalized Kaluza--Klein, gauge gravity or (super)string theories \cite
{v1,v2,v3,vg}.

The concept of N--connection was firstly applied in the framework of Finsler
and Lagrange geometry and gravity and their higher order extensions (see
Refs. \cite{ma,fin,v1,v2} on details). Here it is to be emphasized that
every generalized Finsler like geometry could be modelled equivalently on
higher dimensions (pseudo)Riemannian spacetimes (for some models by
introducing additional torsion and/or nonmetricity structures) with
correspondingly adapted anholonomic frame structures. If we restrict our
considerations only to the higher dimensional Einstein gravity, the induced
N--connection structure becomes a ''pure'' anholonomic frame effect which
points to the fact that the set of dynamical gravitational field variables
given by metric's components was redefined by introducing local frame
variables. The components of a fixed basis define a system (equivalently,
frame) of reference (in four dimensions one uses, for instance, the terms of
vierbien, or tetradic field) with respect to which, in its turn, one states
the coefficients of curved spacetime's metric and of fundamental physical
values and field equations. It should be noted that the procedure of
choosing (establishing) of a system of reference must be also physically
motivated and that in general relativity this task is not considered as a
dynamical one following from the field equations. For trivial models of
physical interactions on curved spacetimes one can restrict our
considerations only with holonomic frames which locally are linearly
equivalent to some coordinate basis. Extending the class of physical fields
and interactions (even in the framework of Einstein's gravity), for example,
by introducing spinor fields, statistical and fluid models with spinning
particles we have to apply frame bundle methods and deal with general
anholonomic frames and phase spaces provided with Cartan's Finsler like
connections (induced by the Levi Civita connection on (pseudo)Riemannian
spacetime) \cite{cartan}.

The modelling of kinetic processes with respect to anholonomic frames (which
induces corresponding N--connection structures) is very useful with the aim
to elucidate flows of fluids of particles not being in local equilibrium.
More exactly, such generic anisotropic fluids are considered to be like in a
thermodynamic equilibrium with respect to some frames which are locally
adapted both to the spacetime metric and N--connection structures but in
general (for instance, with respect to local coordinate frames) the
conditions of local equilibrium are not satisfied.

We developed a proper concept of locally anisotropic spacetime (in brief,
la--spacetime) \cite{v1,v2,v5} in order to provide a unified (super)
geometric background for generalized Kaluza--Klein (super) gravities and low
energy (super) string models when the higher dimension spacetime is
characterized by generic local anisotropies and compatible metric,
N--connection and correspondingly adapted linear connection structures. It
should be noted that the term la--spacetime can be used even for Einstein
spaces if they are provided with anholonomic frame structures. Our treatment
of local anisotropy is more general than that from Refs. \cite{bogoslovsky}
which is used for a subclass of Finsler like metrics being conformally
equivalent (with conformal factors depending both on spacetime coordinates
and tangent vectors) to the flat (pseudo) Euclidean, equivalently,
Minkovschi metric. If in the Bogoslovsky, Goenner and Asanov works \cite
{bogoslovsky,fin} on Finsler like spacetimes there are investigated possible
effects of violation of local Lorentz and/or Poincar\'e symmetries, our
approach to locally anisotropic strings, field interactions and stochastics
\cite{v1,v2,v5} is backgrounded on the fact that such models could be
constructed as to be locally Lorentz invariant with respect to frames
locally adapted to N--connection structures.

The purpose of the present article is to formulate a Boltzmann type kinetic
theory and non--equilibrium ther\-mo\-dy\-na\-mics on spacetimes of higher
dimension with respect to anholonomic frames which models local
an\-isot\-ro\-pi\-es in both type Einstein or generalized
Fins\-ler--Ka\-luza--Klein theories. We shall also discuss possible
applications in modern cosmology and astrophysics.

On kinetic and thermodynamic part our notations and derivations are often
inspired by the Refs. \cite{boisseau} and \cite{groot2}. A number of
formulas will be similar for both locally isotropic and anistropic
spacetimes if in the last case we shall consider the equations and vector,
tensor, spinor and connection objects with respect to anholonomic rest
frames locally adapted to the N--connection structures. In consequence, some
tedious proofs and intermediary formulas will be omitted by referring the
reader to the corresponding works.

To keep the article self--consistent we start up in Section 2 with an
overview of the so--called locally anisotropic spacetime geometry and
gravity. In Sections 3 and 4 we present the basic definitions for locally
anisotropic distribution functions, particle flow and energy momentum
tensors and generalize the kinetic equations for la--spacetimes. The
equilibrium state and derivation of expressions for the particle density and
the entropy density for systems that are closed to a locally anisotropic
state of chemical equilibrium are considered in Section 5. Section 6 is
devoted to the linearized locally anisotropic transport theory. There we
discuss the problem of solution of kinetic equations, prove the linear lows
for locally anisotropic non--equilibrium thermodynamics and obtain the
explicit formulas for transport coefficients. As an example, in Section 7,
we examine the transport theory in curved spaces with rotation ellipsoidal
horizons. Concluding remarks are contained in Section 8.

\section{Spacetimes with Local Anisotropy}

We outline the necessary background on anholonomic frames and nonlinear
connections (N--connection) modelling local anisotropies (la) in curved
spaces and on locally anisotropic gravity \cite{v1,v2} (see Refs. \cite{haw}
and \cite{ma} for details on spacetime differential geometry and
N--connections structures). We shall prove that the Cartan's moving frame
method \cite{cartan1} allows a geometric treatment of both type of locally
isotropic (for simplicity, we shall consider (pseudo) Riemannian spaces) and
anisotropic (the so--called generalized Finsler--Kaluza--Klein spaces).

\subsection{Anholonomic frames and Einstein equations}

In this paper spacetimes are modelled as smooth (i.e class $C^\infty )$
manifolds $V_{(d)}$ of finite integer dimension $d\geq 3,4,...,$ being
Hausdorff, paracompact and connected. We denote the local coordinates on $%
V_{(d)}$ by variables $u^\alpha ,$ where Greek indices $\alpha ,\beta
,...=3,4,...$ could be both type coordinate or abstract (Penrose's) ones. A
spacetime is provided with corresponding geometric structures of symmetric
metric $g_{\alpha \beta }$ and of linear, in general nonsymmetric,
connection $\Gamma _{~\beta \gamma }^\alpha $ defining the covariant
derivation $\nabla _\alpha $ satisfying the metricity conditions $\nabla
_\alpha g_{\beta \gamma }=0.$ We shall underline indices, $\underline{\alpha
},\underline{\beta },...,$ if one would like to emphasize them as abstract
ones.

Let a set of basis (frame) vectors $e^{\underline{\alpha }}=\{e_\alpha ^{%
\underline{\alpha }}=g_{\alpha \beta }e^{\underline{\alpha }\beta }\}$ on $%
V_{(d)}$ be numbered by an underlined index. We shall consider only frames
associated to symmetric metric structures via relations of type
$$
e_{\underline{\alpha }}^\alpha e_{\underline{\beta }\alpha }=\eta _{%
\underline{\alpha },\underline{\beta }}%
\mbox{ and }
e_\alpha ^{\underline{\alpha }} e_\beta ^{\underline{\beta }}\eta _{%
\underline{\alpha },\underline{\beta }}=g_{\alpha \beta },
$$
where the Einstein summation rule is accepted and $\eta _{\underline{\alpha }%
,\underline{\beta }}$ is a given constant symmetric matrix, for simplicity a
pseudo--Euclidean metric of signature $\left( -,+....+\right) $ (the sign
minus is used in this work for the time like coordinate of spacetime).
Operations with underlined and non--underlined indices are correspondingly
performed by using the matrix $\eta _{\underline{\alpha },\underline{\beta }%
},$ its inverse $\eta ^{\underline{\alpha }\underline{\beta }},$ and the
metric $g_{\alpha \beta }$ and its inverse $g^{\alpha \beta }.$ A frame
(local basis) structure $e_\alpha $ on $V_{(d)}$ is characterized by its
anholonomy coefficients $w_{~\beta \gamma }^\alpha $ defined from relations
\begin{equation}
\label{anholon}e_\alpha e_\beta -e_\beta e_\alpha =w_{~\alpha \beta }^\gamma
e_\gamma .
\end{equation}

With respect to a fixed basis $e_\alpha $ and its dual $e^\beta $ we can
decompose tensors and write down their components, for instance,%
$$
T=T_{~\alpha \beta }^\gamma ~e_\gamma \otimes e^\alpha \otimes e^\beta
$$
where by $\otimes $ it is denoted the tensor product.

A spacetime $V_{(d)}$ is {\bf \ holonomic (locally integrable)} if it admits
a frame structure for which the anholonomy coefficients from (\ref{anholon})
vanishes, i.e. $w_{~\alpha \beta }^\gamma =0.$ In this case we can introduce
local coordinate bases,
\begin{equation}
\label{pder}\partial _\alpha =\partial /\partial u^\alpha
\end{equation}
and their duals
\begin{equation}
\label{pdif}d^\alpha =du^\alpha
\end{equation}
and consider components of geometrical objects with respect to such frames.

We note that the general relativity theory was formally defined on holonomic
pseudo-Riemannian manifolds. Even on holonomic spacetimes, for various
(geometrical, computational and physically motivated) purposes, it is
convenient to use anholonomic frames $e_{\underline{\alpha }},$ but we
emphasize that for such spacetimes one can always define some linear
transforms of frames to a coordinate basis, $e_{\underline{\alpha }}=a_{%
\underline{\alpha }}^{\underline{\alpha ^{\prime }}}\partial _{\underline{%
\alpha }^{\prime }}$. By applying both holonomic and anholonomic frames and
theirs mutual transforms on holonomic pseudo-Riemannian spaces there were
developed different variants of tetradic and spinor gravity and extensions
to linear, affine and de Sitter gauge group gravity models \cite
{gaugegr,pr,fock}.

A spacetime is generically{\bf \ anholonomic (locally non-integrable)} if it
does not admit a frame structure for which the anholonomy coefficients from (%
\ref{anholon}) vanishes, i.e. $w_{~\alpha \beta }^\gamma \neq 0.$ In this
case the anholonomy becomes a proper spacetime characteristics. For
instance, a generic anholonomy could be obtained if we consider nontrivial
reductions from higher dimension spaces to lower dimension ones. It induces
nonvanishing additional terms into the torsion,
$$
T\left( \delta _\gamma ,\delta _\beta \right) =T_{~\beta \gamma }^\alpha
\delta _\alpha ,
$$
and curvature,
$$
R\left( \delta _\tau ,\delta _\gamma \right) \delta _\beta =R_{\beta ~\gamma
\tau }^{~\alpha }\delta _\alpha ,
$$
tensors of a linear connection $\Gamma _{~\beta \gamma }^\alpha ,$ with
coefficients defined respectively as
\begin{equation}
\label{torsion}T_{~\beta \gamma }^\alpha =\Gamma _{~\beta \gamma }^\alpha
-\Gamma _{~\gamma \beta }^\alpha +w_{~\beta \gamma }^\alpha
\end{equation}
and
\begin{equation} \label{curvature}
R_{\beta ~\gamma \tau }^{~\alpha } =  \delta _\tau
\Gamma _{~\beta \gamma }^\alpha -
 \delta _\gamma \Gamma _{~\beta \delta }^\alpha +
 \Gamma _{~\beta \gamma }^\varphi \Gamma _{~\varphi \tau }^\alpha
-\Gamma _{~\beta \tau }^\varphi \Gamma _{~\varphi \gamma }^\alpha +
\Gamma _{~\beta
\varphi }^\alpha w_{~\gamma \tau }^\varphi . 
\end{equation}

The Ricci tensor is defined
\begin{equation}
\label{ricci}R_{\beta \gamma }=R_{\beta ~\gamma \alpha }^{~\alpha }
\end{equation}
and the scalar curvature is
\begin{equation}
\label{scalarcurvature}R=g^{\beta \gamma }R_{\beta \gamma }.
\end{equation}

The Einstein equations on an anholonomic spacetime are introduced in a
standard manner,
\begin{equation}
\label{einsteq1}R_{\beta \gamma }-\frac 12g_{\beta \gamma }R=k\Upsilon
_{\beta \gamma },
\end{equation}
where the energy--momentum d--tensor $\Upsilon _{\beta \gamma }$ includes
the cosmological constant terms and possible contributions of torsion (\ref
{torsion}) and matter and $k$ is the coupling constant. For a symmetric
linear connection the torsion field can be considered as induced by some
anholonomy (or equivalently, by some imposed constraints) conditions. For
dynamical torsions there are necessary additional field equations, see, for
instance, the case of locally anisotropic gauge like theories \cite{vg}.

The usual locally isotropic Einstein gravity is obtained on the supposition
that for every anholonomic frame could be defined corresponding linear
transforms to a coordinate frame basis.

It is a topic of further theoretical and experimental investigations to
establish if the present day experimental data on anisotropic structure of
Universe is a consequence of matter and quantum fluctuation induced
anisotropies and for some scales the anisotropy is a consequence of
anholonomy of observer's frame. The spacetime anisotropy could be also a
generic property following, for instance, from string theory, and from a
more general self--consistent gravitational theory when both the left
(geometric) and right (matter energy--momentum tensor) parts of Einstein
equations depend on anisotropic parameters.

\subsection{The local anisotropy and nonlinear connection}

A subclass of anholonomic spacetimes consists from those with local
anisotropy modelled by a nonlinear connection structure. In this subsection
we briefly outline the geometry of anholonomic frames with induced nonlinear
connection structure.

The la--spacetime dimension is split locally into two components, $n$ for
isotropic coordinates and $m$ for an\-isot\-ropic coordinates, when $%
n_{(a)}=n+m$ with $n\geq 2$ and $m\geq 1.$ We shall use local coordinates $%
u^\alpha =(x^i,y^a),$ where Greek indices $\alpha ,\beta ,...$ take values $%
1,2,...,n+m$ and Latin indices $i$ and $a$ are correspondingly $n$ and $m$
dimensional, i.e. $i,j,k...=1,2,$ $...,n$ and $a,b,c,...=1,2,...,m.$

Now, we consider an invariant geometric definition of spacetime's splitting
into isotropic and anisotropic components. For modelling la--spacetimes one
uses a vector bundle ${\cal E}=(E_{n+m},p,M_{(n)},F_{(m)},Gr)$ provided with
{\bf nonlinear connection} (in brief, {\bf N--connection)} structure $%
N=\{N_j^a\left( u^\alpha \right) \},$ where $N_j^a\left( u^\alpha \right) $
are its coefficients \cite{ma}. We use denotations:\ $E_{m+n}$ is a $(n+m)$%
--dimensional total space of a vector bundle, $M_{(n)}$ is the $n$%
--dimensional base manifold, $F_{(m)}$ is the typical fiber being a $m$%
--dimensional real vector space, $Gr$ is the group of automorphisms of $%
F_{(m)}$ and $p$ is a surjective map. For simplicity, we shall consider only
local constructions on vector bundles.

The N--connection is a new geometric object which generalizes that of linear
connection. This concept came from Finsler geometry (see the Cartan's
monograph \cite{cartan}), the global formulation of it is due to W. Barthel
\cite{barthel}, and it is studied in details in Miron and Anastasiei works
\cite{ma}. We have extended the geometric constructions for spinor bundles
and superbundles with further applications in locally anisotropic field
theory and strings and modern cosmology and astrophysics \cite{v1,v2,v3}. We
have also illustrated \cite{v6} that the N--connection could be introduced on
(pseudo) Riemnannian spacetimes and in Einstein gravity if we consider
anholonomic frames consisting from sets of basic vectors some of them being
holonomic and the rest anholonomic. In this case the N--connection
coefficients are associated to the frame structure and transforms into some
metric components if the considerations are transferred with respect to a
coordinate basis.

The rigorous mathematical definition of N--connection is based on the
formalism of horizontal and vertical subbundles and on exact sequences of
vector bundles. Here, for simplicity, we define a N--connection as a
distribution which for every point $u=(x,y)\in {\cal E}$ defines a local
decomposition of the tangent space of our vector bundle, $T_uE,$ into
horizontal, $H_uE,$ and vertical (anisotropy), $V_uE,$ subspaces, i.e.%
$$
T_uE=H_uE\oplus V_uE.
$$

If a N--connection with coefficients $N_j^a\left( u^\alpha \right) $ is
introduced on the vector bundle ${\cal E}$ the modelled spacetime posses a
generic local anisotropy and in this case we can not apply in a usual manner
the operators of partial derivatives and their duals, differentials. Instead
of coordinate bases (\ref{pder}) and (\ref{pdif}) we must consider some
bases adapted to the N--connection structure:
\begin{equation} \label{dder}
\delta _\alpha  = (\delta _i,\partial _a ) =
 \frac \delta {\partial u^\alpha }    = 
 \left( \delta _i = \frac \delta {\partial x^i} =
\frac \partial {\partial x^i} -
N_i^b \left( x^j,y^c\right) \frac \partial {\partial y^b},
 \partial _a  = \frac \partial {\partial y^a}\right)
\end{equation}
and
\begin{equation} \label{ddif}
\delta ^\beta  =  \left( d^i,\delta ^a \right)
  =  \delta u^\beta  = 
\left( d^i = dx^i, \delta ^a = 
\delta y^a =dy^a +N_k^a \left( x^j,y^b \right) dx^k\right) .
\end{equation}

A nonlinear connection (N--connection) is characterized by its curvature
\begin{equation}
\label{ncurv}\Omega _{ij}^a=\frac{\partial N_i^a}{\partial x^j}-\frac{%
\partial N_j^a}{\partial x^i}+N_i^b\frac{\partial N_j^a}{\partial y^b}-N_j^b%
\frac{\partial N_i^a}{\partial y^b}.
\end{equation}
Here we note that the class of usual linear connections can be considered as
a particular case when
$$
N_j^a(x,y)=\Gamma _{bj}^a(x)y^b.
$$

The elongation (by N--connection) of partial derivatives in the adapted to
the N--connection partial derivatives (\ref{dder}), or the locally adapted
basis (la--basis) $\delta _\beta $, reflects the fact that the spacetime $%
{\cal E}$ is locally anisotropic and generically anholonomic because there
are satisfied anholonomy relations (\ref{anholon}),
$$
\delta _\alpha \delta _\beta -\delta _\beta \delta _\alpha =w_{~\alpha \beta
}^\gamma \delta _\gamma ,%
$$
where anolonomy coefficients are as follows
\begin{eqnarray}
w_{~ij}^k & = & 0,w_{~aj}^k=0,w_{~ia}^k=0,w_{~ab}^k=0,w_{~ab}^c=0,
\nonumber\\
w_{~ij}^a & = &
-\Omega _{ij}^a,w_{~aj}^b=-\partial _aN_i^b,w_{~ia}^b=\partial _aN_i^b.
\nonumber
\end{eqnarray}

On a la--spacetime the geometrical objects have a distinguished (by
N--connec\-ti\-on), into horizontal and vertical components, character. They
are briefly called d--tensors, d--metrics and/or d--connections. Their
components are defined with respect to a la--basis of type (\ref{dder}), its
dual (\ref{ddif}), or their tensor products (d--linear or d--affine
transforms of such frames could also be considered). For instance a
covariant and contravariant d--tensor $Z,$ is expressed as%
\begin{equation}
Z =  Z_{~\beta }^\alpha \delta _\alpha \otimes \delta ^\beta 
 =  Z_{~j}^i\delta _i\otimes d^j+Z_{~a}^i\delta _i\otimes \delta^a+ 
Z_{~j}^b\partial _b\otimes d^j+Z_{~a}^b\partial _b\otimes \delta ^a.
\end{equation}

A symmetric d--metric on la--space ${\cal E}$ is written as
\begin{equation} \label{dmetric}
\delta s^2  =  g_{\alpha \beta }\left( u\right)
  \delta ^\alpha \otimes \delta^\beta = 
 g_{ij}(x,y)dx^idx^j+h_{ab}(x,y)\delta y^a\delta y^b. \nonumber
\end{equation}

A linear d--connection $D$ on la--space ${\cal E,}$
$$
D_{\delta _\gamma }\delta _\beta =\Gamma _{~\beta \gamma }^\alpha \left(
x^k,y\right) \delta _\alpha ,%
$$
is parametrized by non--trivial h--v--components,
\begin{equation}
\label{dcon}\Gamma _{~\beta \gamma }^\alpha =\left(
L_{~jk}^i,L_{~bk}^a,C_{~jc}^i,C_{~bc}^a\right) .
\end{equation}
Some d--connection and d--metric structures are compatible if there are
satisfied the conditions
$$
D_\alpha g_{\beta \gamma }=0.%
$$
For instance, a canonical compatible d--connection
$$
^c\Gamma _{~\beta \gamma }^\alpha =\left(
^cL_{~jk}^i,^cL_{~bk}^a,^cC_{~jc}^i,^cC_{~bc}^a\right)
$$
is defined by the coefficients of d--metric (\ref{dmetric}), $g_{ij}\left(
x,y\right) $ and $h_{ab}\left( x,y\right) ,$ and by the coefficients of
N--connection,%
\begin{eqnarray}
^cL_{~jk}^i & = & \frac 12g^{in}\left( \delta _kg_{nj}+\delta _jg_{nk}-\delta
_ng_{jk}\right) , \label{cdcon} \\
^cL_{~bk}^a & = & \partial _bN_k^a+\frac 12h^{ac}\left( \delta
_kh_{bc}-h_{dc}\partial _bN_i^d-h_{db}\partial _cN_i^d\right) ,
\nonumber \\
^cC_{~jc}^i & = & \frac 12g^{ik}\partial _cg_{jk}, \nonumber \\
^cC_{~bc}^a & = & \frac 12h^{ad}\left( \partial _ch_{db}+\partial
_bh_{dc}-\partial _dh_{bc}\right)  \nonumber
\end{eqnarray}
The coefficients of the canonical d--connection generalize for
la--spacetimes the well known Cristoffel symbols.

For a d--connection (\ref{dcon}) we can compute the components of, in our
case d--torsion, (\ref{torsion})
\begin{eqnarray}
T_{.jk}^i & = & T_{jk}^i=L_{jk}^i-L_{kj}^i,\quad
T_{ja}^i=C_{.ja}^i,T_{aj}^i=-C_{ja}^i, \nonumber \\
T_{.ja}^i & = & 0,\quad T_{.bc}^a=S_{.bc}^a=C_{bc}^a-C_{cb}^a,
\nonumber \\
T_{.ij}^a & = &
-\Omega _{ij}^a,\quad T_{.bi}^a= \partial _b  N_i^a
-L_{.bj}^a,\quad T_{.ib}^a=-T_{.bi}^a. \nonumber
\end{eqnarray}

In a similar manner, putting non--vanishing coefficients (\ref{dcon}) into
the formula for curvature (\ref{curvature}), we can compute the non--trivial
components of a d--curvature
\begin{eqnarray}
R_{h.jk}^{.i} & = & \delta _kL_{.hj}^i-\delta_jL_{.hk}^i 
 +  L_{.hj}^mL_{mk}^i-L_{.hk}^mL_{mj}^i-C_{.ha}^i\Omega _{.jk}^a,
\nonumber \\
R_{b.jk}^{.a} & = & \delta _kL_{.bj}^a-\delta_jL_{.bk}^a 
  +  L_{.bj}^cL_{.ck}^a-L_{.bk}^cL_{.cj}^a-C_{.bc}^a\Omega _{.jk}^c,
\nonumber \\
P_{j.ka}^{.i} & = & \partial _kL_{.jk}^i +C_{.jb}^iT_{.ka}^b 
 -  ( \partial _kC_{.ja}^i+L_{.lk}^iC_{.ja}^l -
L_{.jk}^lC_{.la}^i-L_{.ak}^cC_{.jc}^i ), \nonumber \\
P_{b.ka}^{.c} & = & \partial _aL_{.bk}^c +C_{.bd}^cT_{.ka}^d 
  - ( \partial _kC_{.ba}^c+L_{.dk}^{c\,}C_{.ba}^d
- L_{.bk}^dC_{.da}^c-L_{.ak}^dC_{.bd}^c ) \nonumber \\
S_{j.bc}^{.i} & = & \partial _cC_{.jb}^i-\partial _bC_{.jc}^i
 +  C_{.jb}^hC_{.hc}^i-C_{.jc}^hC_{hb}^i, \nonumber \\
S_{b.cd}^{.a} & = &\partial _dC_{.bc}^a-\partial
_cC_{.bd}^a+C_{.bc}^eC_{.ed}^a-C_{.bd}^eC_{.ec}^a. \nonumber
\end{eqnarray}

The components of the Ricci tensor (\ref{dricci}) with respect to locally
adapted frames (\ref{dder}) and (\ref{ddif}) (in this case, d--tensor) are
as follows:%
\begin{eqnarray}
R_{ij} & = & R_{i.jk}^{.k},\quad
 R_{ia}=-^2P_{ia}=-P_{i.ka}^{.k},\label{dricci} \\
R_{ai} &= & ^1P_{ai}=P_{a.ib}^{.b},\quad R_{ab}=S_{a.bc}^{.c}. \nonumber
\end{eqnarray}
We point out that because, in general, $^1P_{ai}\neq ~^2P_{ia}$ the Ricci
d-tensor is non symmetric.

Having defined a d-metric of type (\ref{dmetric}) in ${\cal E}$ we can
compute the scalar curvature (\ref{scalarcurvature}) of a d-connection $D,$%
\begin{equation}
\label{dscalar}{\overleftarrow{R}}=G^{\alpha \beta }R_{\alpha \beta }=%
\widehat{R}+S,
\end{equation}
where $\widehat{R}=g^{ij}R_{ij}$ and $S=h^{ab}S_{ab}.$

Now, by introducing the values (\ref{dricci}) and (\ref{dscalar}) into
anholonomic gravity field equations (\ref{einsteq1}) we can write down the
system of Einstein equations for la--gravity with prescribed N--connection
structure \cite{ma}:%
\begin{eqnarray}
R_{ij}-\frac 12\left( \widehat{R}+S\right) g_{ij} & = &
k\Upsilon _{ij}, \label{einsteq2} \\
S_{ab}-\frac 12\left( \widehat{R}+S\right) h_{ab} & = & k\Upsilon _{ab},
\nonumber \\
^1P_{ai} & = & k\Upsilon _{ai}, \nonumber \\
^2P_{ia} & = & -k\Upsilon _{ia}, \nonumber
\end{eqnarray}
where $\Upsilon _{ij},\Upsilon _{ab},\Upsilon _{ai}$ and $\Upsilon _{ia}$
are the components of the energy--momentum d--tensor field. We note that
such decompositions into h-- and v--components of gravitational field
equations have to be considered even in general relativity if physical
interactions are examined with respect to an anholonomic frame of reference
with associated N--connection structure.

There are variants of la--gravitational field equations derived in the
low--energy limits of the theory of locally anisotropic (super)strings \cite
{v2} or in the framework of gauge like la--gravity \cite{vg,v3} when the
N--connection and torsions are dynamical fields and satisfy some additional
field equations.

\subsection{Modelling of generalized Finsler geometries in \protect\newline %
(pseudo) Riemannian spaces}

The present day trend is to consider the Finsler like geometries and their
generalizations as to be quite sophisticate for straightforward applications
in quantum and classical field theory. The aim of this subsection is to
proof that, a matter of principle, such geometries could be equivalently
modelled on corresponding (pseudo) Riemannian manifolds (tangent or vector
bundles) by using the Cartan's moving frame method and from this viewpoint a
wide class of Finsler like metrics could be treated as some solutions of
usual Einstein field equations.

\subsubsection{Almost Hermitian Models of Lagrange and Finsler Spaces}

This topic was originally investigated by Miron and Anastasiei \cite{ma};
here we outline some basic results.

Let us model a la--spacetime not on a vector bundle ${\cal E}$ but on a
manifold $\widetilde{TM}=TM\backslash \{0\}$ associated to a tangent bundle $%
TM$ of a $n$--dimensional base space $M$ (when the dimensions of the typical
fiber and base are equal, $n=m$ and $\backslash \{0\}$ means that there is
eliminated the null cross--section of the bundle projection $\tau
:TM\rightarrow M)$ and consider d--metrics of type
\begin{equation} \label{dmetricl}
\delta s^2  =  g_{\alpha \beta }\left( u\right)
  \delta ^\alpha \otimes \delta^\beta 
 =  g_{ij}(x,y)dx^idx^j+g_{ij}(x,y)\delta y^i\delta y^i .\nonumber
\end{equation}
On $TM$ we can define a natural almost complex structure $C_{(a)}$ as
follows
$$
C_{(a)}\left( \delta _i\right) =-\partial /\partial y^i
 \mbox{ and } C_{(a)}\left( \partial /\partial y^i\right) =\delta _i
$$
where the la--derivative $\delta _i=\partial /\partial x^i-N_i^k\partial
/\partial y^k$ (\ref{dder}) and la--differential $\delta ^i=dy^i+N_k^idx^k$ (%
\ref{ddif}) act on $\widetilde{TM}$ being adapted to a nontrivial
N--connection structure $N=\{N_j^k\left( x,y\right) \}$ in $TM$. It is
obvious that $C_{(a)}^2=-I.$ The pair $\left( \delta s^2,C_{(a)}\right) $
defines an almost Hermitian structure on $\widetilde{TM}$ with an associate
2--form%
$$
\theta =g_{ij}\left( x,y\right) \delta ^i\Lambda dx^j
$$
and the triad $K^{2n}=\left( \widetilde{TM},\delta s^2,C_{(a)}\right) $ is
an almost K\"ah\-le\-ri\-an space. By straightforward calculations we can
verify that the canonical d--connection (\ref{dcon}) satisfies the
conditions
$$
~^cD_X\left( \delta s^2\right) =0,~^cD_X\left( C_{(a)}\right) =0
$$
for any d--vector $X$ on $TM$ and has zero $hhh$-- and $vvv$--torsions.

The notion of {\bf Lagrange space} \cite{kern,ma} was introduced as a
generalization of Finsler geometry in order to geometrize the fundamental
concepts in mechanics. A regular Lagrangian $L\left( x^i,y^i\right) $ on $%
\widetilde{TM}$ is introduced as a continuity class $C^\infty $ function $%
L:TM\rightarrow \R$ for which the matrix
\begin{equation}
\label{lagrm}g_{ij}\left( x,y\right) =\frac 12\frac{\partial ^2L}{\partial
y^i\partial y^j}
\end{equation}
has rank $n$ and is of constant signature on $\widetilde{TM}.$ A d--metric (%
\ref{dmetricl}) with coefficients of form (\ref{lagrm}), a corresponding
canonical d--connection (\ref{dcon}) and almost complex structure $C_{(a)}$
defines an almost Hermitian model of Lagrange geometry.

For arbitrary metrics $g_{ij}\left( x,y\right) $ of rank $n$ and constant
signature on $\widetilde{TM}$, which can not be determined as a second
derivative of a Lagrangian, one defines the so--called generalized Lagrange
geometry on $TM$ (see details in \cite{ma}).

A particular subclass of metrics of type (\ref{lagrm}) consists from those
where instead of a regular Lagrangian one considers a Finsler metric
function $F$ on $M$ defined as $F:TM\rightarrow \R$ having the properties
that it is of class $C^\infty $ on $\widetilde{TM}$ and only continuous on
the image of the null cross--section in $TM,$ the restriction of $F$ on $%
\widetilde{TM}$ is a positive function homogeneous of degree 1 with respect
to the variables $y^i,$ i. e.
$$
F\left( x,\lambda y\right) =\lambda F\left( x,y\right) ,\lambda \in \R^n,
$$
and the quadratic form on $\R^n$ with coefficients
\begin{equation}
\label{finm}g_{ij}\left( x,y\right) =\frac 12\frac{\partial ^2F^2}{\partial
y^i\partial y^j},
\end{equation}
defined on $\widetilde{TM,}$ is positive definite. Different approaches to
Finsler geometry, its generalizations and applications are examined in a
number of monographs \cite{fin,cartan,ma} and as a rule they are based on
the assertion that in this type of geometries the usual (pseudo)Riemannian
metric interval
$$
ds=\sqrt{g_{ij}\left( x\right) dx^idx^j}
$$
on a manifold $M$ is changed into a nonlinear one defined by the Finsler
metric $F$ (fundamental function) on $\widetilde{TM}$ (we note an ambiguity
in terminology used in monographs on Finsler geometry and on gravity
theories with respect to such terms as Minkowschi space, metric function and
so on)
\begin{equation}
\label{finint}ds=F\left( x^i,dx^j\right) .
\end{equation}
Geometric spaces with a 'combersome' variational calculus and a number of
curvatures, torsions and invariants connected with nonlinear metric
intervals of type (\ref{lagrm}) are considered as less suitable for purposes
of modern field and particle physics.

In our investigations of generalized Finsler geometries in (super) string,
gravity and gauge theories \cite{v1,v2} we advocated the idea that instead
of usual geometric constructions based on straightforward applications of
derivatives of (\ref{finm}) following from a nonlinear interval (\ref{finint}%
) one should consider d--metrics (\ref{dmetric}) and/or (\ref{dmetricl})
with the coefficients of necessity determined via an almost Hermitian model
of a Lagrange (\ref{lagrm}), Finsler geometry (\ref{finm}) and/or their
extended variants. This way, by a synthesis of the moving frame method with
the geometry of N--connection, we can investigate a various class of higher
and lower dimension gravitational models with generic or induced
anisotropies in a unified manner on some anholonomic and/or Kaluza--Klein
spacetimes.

As a matter of principle, having a physical model with a d--metric and
geometrical objects associated to la--frames, we can redefine the physical
values with respect to a local coordinate base on a (pseudo) Riemannian
space. The coefficients $g_{ij}\left( x,y\right) $ and $h_{ab}\left(
x,y\right) $ of d--metric written for a la--basis $\delta u^\alpha =\left(
dx^i,\delta y^a\right) $ transforms into a usual (pseudo) Riemannian metric
if we rearrange the components with respect to a local coordinate basis $du^{%
\widehat{\alpha }}=\left( dx^{\widehat{i}},dy^{\widehat{a}}\right) $,
\begin{equation}
\label{kkm}g_{\widehat{\alpha }\widehat{\beta }}=\left(
\begin{array}{cc}
g_{\widehat{i}\widehat{j}}+N_{\widehat{i}}^{\widehat{a}}N_{\widehat{j}}^{%
\widehat{b}}h_{\widehat{a}\widehat{b}} & N_{
\widehat{j}}^{\widehat{e}}h_{\widehat{a}\widehat{e}} \\ N_{\widehat{i}}^{%
\widehat{e}}h_{\widehat{b}\widehat{e}} & h_{\widehat{a}\widehat{b}}
\end{array}
\right),
\end{equation}
where 'hats' on indices emphasize that coefficients of the metric are given
with respect to a coordinate (holonomic) basis on a spacetime $V^{n+m}.$
Parametrizations (ansatzs) of metrics of type (\ref{kkm}) are largely
applied in Kaluza--Klein gravity and its generalizations \cite{over}. In our
works \cite{v1,v2,v3,v6}, following the geometric constructions from \cite
{ma}, we proved that the physical model of interactions is substantially
simplified, as well we can correctly elucidate anisotropic effects, if we
work with diagonal blocks of d--metrics (\ref{dmetric}) with respect to
anholonomic frames determined by a N--connection structure.

\subsubsection{Finsler like metrics in Einstein's gravity}

There are obtained \cite{v3,v6} (see Appendix) some classes of locally
anisotropic cosmological and black hole like solutions (in three, four and
higher dimensions) which can be treated as generalized Finsler metrics being
of nonspheric symmetry (with rotation ellipsoid, torus and cylindrical event
horizons, or with elliptical oscillations of horizons). Under corresponding
conditions such metrics could be solutions of field equations in general
relativity or its lower or higher dimension variants. Here we shall
formulate the general criteria when a Finsler like metric could be a
solution of gravitational field equations in Einstein gravity.

Let consider on $\widetilde{TM}$ an ansatz of type (\ref{kkm}) when $%
g_{ij}=h_{ij}=\frac 12\partial ^2F^2/\partial y^i\partial y^j$ (for
simplicity, we omit 'hats' on indices) i.e.
\begin{equation}
\label{fm}g_{\widehat{\alpha }\widehat{\beta }}=\frac 12\left(
\begin{array}{cc}
\frac{\partial ^2F^2}{\partial y^i\partial y^j}+N_i^kN_j^l\frac{\partial
^2F^2}{\partial y^k\partial y^l} & N_j^l
\frac{\partial ^2F^2}{\partial y^k\partial y^l} \\ N_i^k\frac{\partial ^2F^2%
}{\partial y^k\partial y^l} & \frac{\partial ^2F^2}{\partial y^i\partial y^j}
\end{array}
\right) .
\end{equation}
A metric (\ref{fm}), induced by a Finsler quadratic form (\ref{finm}) could
be treated in a framework of a Kaluza--Klein model if for some values of
Finsler metric $F\left( x,y\right) $ and N--connection coefficients $%
N_i^k\left( x,y\right) $ this metric is a solution of the Einstein equations
(\ref{einsteq1}) written with respect to a holonomic frame. For the
dimension $n=2,$ when the values $F$ and $N$ are chosen to induce locally a
(pseudo) Riemannian metric $g_{\widehat{\alpha }\widehat{\beta }}$ of
signature $\left( -,+,+,+\right) $ and with coefficients satisfying the four
dimensional Einstein equations we define a subclass of Finsler metrics in
the framework of general relativity. Here we note that, in general, a
N--connection, on a Finsler space, subjected to the condition that the
induced (pseudo) Riemannian metric is a solution of usual Einstein equations
does not coincide with the well known Cartan's N--connection \cite
{cartan,fin}. We have to examine possible compatible deformations of
N--connection structures \cite{ma}.

Instead of Finsler like quadratic forms we can consider ansatzs of type (\ref
{kkm}) with $g_{ij}$ and $h_{ij}$ induced by a Lagrange quadratic form (\ref
{lagrm}). A general approach to the geometry of spacetimes with generic
local anisotropy can be developed on embeddings into corresponding
Kaluza--Klein theories and adequate modelling of la--interactions with
respect to anholonomic or holonomic frames and associated N--connection
structures.

\section{Collisionless relativistic kinetic equation}

As argued in Section 2, the spacetimes could be of generic local anisotropy
after nontrivial reductions from some higher dimension theories or posses a
local anisotropy induced by anholonomic frame structures even we restrict
our considerations to the general relativity theory. In this line of
particular interest is the formulation of relativistic kinetic theory with
respect to general anholonomic frames and elucidation of locally anisotropic
kinetic and thermodynamic processes.

\subsection{The distribution function and its moments}

We use the relativistic approach to kinetic theory (we refer readers to
monographs \cite{groot2,vlas} for hystory and complete treatment). Let us
consider a simple system consisting from $\varpi$ point particles of mass $%
\widetilde{m}$ in a la--spacetime with a d--metric $g_{\alpha \beta }.$
Every particle is characterized by its coordinates $u_{(l)}^\alpha =\left(
x_{(l)}^i,y_{(l)}^a\right) ,(u^{\overline{1}}=x^1=ct$ is considered the time
like coordinate, for simplicity we put hereafter the light velocity $c=1);$ $%
x_{(l)}^2,x_{(l)}^3,...,x_{(l)}^n$ and $y_{(l)}^1,y_{(l)}^2,...,y_{(l)}^m$
are respectively space like and anisotropy coordinates, where the index $(l)$
enumerates the particles in the system. The particles momenta are denoted by
$p_{\alpha \left( l\right) }=g_{\alpha \beta }p_{(l)}^\beta ,$ . We shall
use the distribution function $\Phi \left( u^\alpha ,p_\beta \right) ,$
given on the space of supporting elements $\left( u^\alpha ,p_\beta \right)
, $ as a general characteristic of particle system.

The system (of particles) is to be defined by using the random function
\begin{equation} \label{rf}
\phi \left( u^\alpha ,p_\beta \right)  = 
\sum\limits_{l=1}^\varpi \int \delta s  \delta ^{(n_{(a)})}
\left( u^\alpha -u_{(l)}^\alpha (s)\right) \ 
\delta ^{(n_{(a)})}\left( p^\beta -p_{(l)}^\beta (s)\right) , \nonumber
\end{equation}
where the sum is taken on all system's particles,
$$
\delta s=\sqrt{g_{\alpha\beta }\delta u^\alpha \delta u^\beta }%
$$
is the interval element along the particle trajectory $u_{(l)}^\alpha (s)$
parametrized by a natural parameter $s$ and $\delta ^{n_{(a)}}\left(
u^\alpha \right) $ is the $n_{(a)}$--dimensional delta function. The
functions $u_{(l)}^\alpha (s)$ and $p_{(l)}^\beta (s)$ describing the
propagation of the $l$--particle are found from the motion equations on
la--spacetime
\begin{eqnarray}
\label{motioneq} &\widetilde{m}&\frac{\delta u_{(l)}^\alpha }{ds} =
p_{(l)}^\alpha  \\
&\widetilde{m}&
D\frac{\delta p_{\alpha (l)}}{ds} = \widetilde{m}\frac{\delta
p_{\alpha (l)}}{ds}-~^c\Gamma _{~\alpha \beta }^\tau \left( u_{(l)}^\gamma
(s)\right) p_{\tau (l)}p_{(l)}^\beta =F_{\alpha (l)}, \nonumber
\end{eqnarray}
where $^c\Gamma _{~\beta \gamma }^\alpha $ is the canonical d--connection
with coefficients (\ref{cdcon}) and by $F_{\alpha (l)}$ is denoted an
exterior force (electromagnetic or another type) acting on the $l$th
particle.

The distribution function $\Phi \left( u,p\right) $ is defined by averaging
on paths of random function
$$
\Phi \left( u,p\right) =<<\phi \left( u,p\right) >>
$$
where brackets $<<...>>$ denote path averaging.

Let consider a space like hypersurface ${\cal F}(u^\alpha )=const$ with
elements $\delta \Sigma _\alpha =n_\alpha \delta \Sigma ,$ where $n_\alpha =%
\frac{D_\alpha {\cal T}}{\left| D{\cal T}\right| }$ and $d\Sigma =\frac{%
\sqrt{|g|}\delta ^{n_{(a)}}u}{n_\alpha \delta u^\alpha },$ $\left| D{\cal T}%
\right| =\sqrt{g^{\alpha \beta }D_\alpha D_\beta {\cal T}}$ and $\sqrt{|g|}%
\delta ^{n_{(a)}}u$ is the invariant space--time volume. A local system of
reference in a point $u_{(0)}^\alpha ,$ with the metric $g_{ij}^{(0)}=\eta
_{ij}=diag\left(- 1,1,1,...,1\right) ,$ is obtained if ${\cal F}%
(u_{(0)}^\alpha )=x_{(0)}^1$ . In this case $\delta \Sigma _\alpha =\delta
_\alpha ^1\delta ^{n_{(a)}-1}u,$ $\delta ^{n_{(a)}-1}u=dx^2dx^3...dx^n\delta
y^1\delta y^2...\delta y^m$ and $\delta _\beta ^\alpha $ is the Kronecker
symbol. The value
$$
\Phi \left( u,p\right) v^\alpha d\Sigma _\alpha \frac{\delta ^{n_{(a)}}p}{%
\sqrt{|g|}},
$$
where $v^\alpha =p^\alpha /\widetilde{m}$ and 
$\delta P=\delta ^{n_{(a)}}p/\sqrt{|g|}=\delta p_1\delta p_2...\delta
p_{n_{(a)}}/\sqrt{|g|}$ 
is the invariant volume in the momentum space and defines the quantity of
particles intersecting the hypersurface element $\delta \Sigma _\alpha $
with momenta $p_\alpha $ included into the element $\delta P$ in the
vicinity of the point $u^\alpha .$ The first $<p^\alpha >$ and second
moments $<p^\alpha p^\beta >$ of distribution function $\Phi \left(
x,p\right) $ give respectively the flux of particles $n^\alpha $ and the
energy--momentum $T^{\alpha \beta },$%
\begin{equation}
\label{fluxpart}<p^\alpha >=\int \Phi \left( u,p\right) p^\alpha \delta P=%
\widetilde{m}n^\alpha
\end{equation}
and
\begin{equation}
\label{fluxen}<p^\alpha p^\beta >=\int \Phi \left( u,p\right) p^\alpha
p^\beta \delta P=\widetilde{m} \Upsilon ^{\alpha \beta }.
\end{equation}

We emphasize that the motion equations (\ref{motioneq}) have the first
integral, $g^{\alpha \beta }p_\alpha ^{(l)}p_\beta ^{(l)}=\widetilde{m}%
^2=const$ for every $l$--particle, so the functions (\ref{rf}) and, in
consequence, $\Phi \left( u,p\right) $ are non--zero only on the mass
hypersurface
\begin{equation}
\label{maseq}g^{\alpha \beta }p_\alpha p_\beta =\widetilde{m}^2,
\end{equation}
which in distinguished by a N--connection form (see the dual to d--metric (%
\ref{dmetric})) is written
$$
g^{ij}p_ip_j+g^{ab}p_ap_b=\widetilde{m}^2.
$$

For computations it is convenient to use a new distribution function $%
f\left( u,p\right) $ on a $n_{(a)} -1$ dimensional la--hyperspace (\ref
{maseq})
$$
\Phi \left( u^\alpha ,p_\beta \right) =f\left( u^\alpha ,p_{\widehat{\beta }%
}\right) \delta \left( \sqrt{g^{\alpha \beta }p_\alpha p_\beta }-\widetilde{m%
}\right) \theta \left( p_{\overline{1}}\right)
$$
where
$$
\theta \left( p_{\overline{1}}\right) =\left\{
\begin{array}{c}
1,~p_{
\overline{1}}\geq 0 \\ 0,~p_{\overline{1}}<0
\end{array}
\right\}
$$
and the index $\widehat{\beta }$ runs values in $n_{(a)}-1$ dimensional
la--space. The flux of particles (\ref{fluxpart}) and of energy--momentum (%
\ref{fluxen}) are computed by using $f(u,p)$ as
$$
n^\alpha =\int \delta \varsigma ~p^\alpha f\left( u^\gamma ,p_{\widehat{%
\beta }}\right) ,
$$
with respective base $n^i$ and fiber $n^a$ components,
$$
n^i=\int \delta \varsigma ~p^if\left( u^\gamma ,p_{\widehat{\beta }}\right)
 \mbox{ and }
n^a=\int \delta \varsigma ~p^af\left( u^\gamma ,p_{\widehat{\beta }}\right) ,%
$$
 when 
$$
\Upsilon ^{\alpha \beta }=\int \delta \varsigma ~p^\alpha p^\beta f\left(
u^\gamma ,p_{\widehat{\beta }}\right)
$$
where $p^{\overline{1}}$ is expressed via $p^2,p^3,...,p^{n_{(a)}}$ by using
the equation (\ref{maseq}) and it is used the abbreviation $\delta \varsigma
=\delta ^{n_{(a)}-1}p/(\sqrt{|g|}p^{\overline{1}}).$ The energy--momentum
can be also split into base--fiber (horizontal--vertical, in brief, h-- and
v--, or hv--components) by a corresponding distinguishing of momenta, $%
\Upsilon ^{\alpha \beta }=\{\Upsilon ^{ij},\Upsilon ^{aj},\Upsilon
^{ib},\Upsilon ^{ab}\}.$

The one--particle distribution function $f\left( u,p\right) $ characterizes
the number of particles (with mass $\widetilde{m},$ or massless if $%
\widetilde{m}=0)$ at a point $u^\alpha $ of a la--spacetime of dimension $%
n_{(a)}$ having distinguished by N--connection momentum vector (in brief,
momentum d--vector) $p^\alpha =(p^i,p^a)=\left( p^{\overline{1}},%
\overrightarrow{p}\right) .$ One states that
$$
f\left( u,p\right) p^\beta ~\delta \sigma _\beta ~\delta \varsigma =f\left(
u,p\right) p^i~\delta \sigma _i~\delta \varsigma +f\left( u,p\right)
p^a~\delta \sigma _a~\delta \varsigma ,
$$
gives the number of world--lines of particles with momentum d--vector $%
p^\alpha $ in an interval $\delta p^\alpha $ around $p,$ crossing a space
like hypersurface $\delta \sigma _\beta $ at a point $u.$ When $\delta
\sigma _\beta $ is taken to be time like one has $\delta \sigma _\beta
=\left( \delta ^{n_{(a)}}u,0,...,0\right) .$

The velocity d--field of a fluid $U^\mu $ in la--spacetime can be defined in
some way like in relativistic kinetic theory \cite{groot2}. To obtain a
particularly simple form for the energy--momentum d--tensor one should
follow the Landau--Lifshitz approach \cite{landau} when the locally
anisotropic fluid velocity $U^\mu \left( u\right) =\left( U^i\left( u\right)
,U^a\left( u\right) \right) $ and the energy density $\varepsilon \left(
u\right) $ are defined respectively as the eigenvector and eigenvalue of the
eigenvalue equation
\begin{equation}
\label{llcond}\Upsilon ^{\alpha \beta }U_\beta =\varepsilon U^\alpha .
\end{equation}
A unique value of $U^\alpha $ is to be found from the conditions to be a
time like and normalized to the unity d--vector, 
 $U^\alpha U_\alpha =1.$ 

Having so fixed the fluid d--velocity we can define correspondingly the
particle density
\begin{equation}
\label{pdens}\widetilde{n}\left( u\right) =n^\alpha U_\alpha ,
\end{equation}
the energy density
\begin{equation}
\label{edens}\widetilde{\varepsilon }\left( u\right) =U_\alpha \Upsilon
^{\alpha \beta }U_\beta
\end{equation}
and the average energy per particle
\begin{equation}
\label{edensp}e=\widetilde{\varepsilon }/\widetilde{n}.
\end{equation}

We can introduce also the pressure d--tensor
\begin{equation}
\label{pressproj}P^{\alpha \beta }\left( u\right) =\Delta ^{\alpha \tau
}\Upsilon _{\tau \nu }\Delta ^{\nu \beta }
\end{equation}
by applying the projector
\begin{equation}
\label{proj}\Delta ^{\alpha \beta }=g^{\alpha \beta }-U^\alpha U^\beta
\end{equation}
with properties%
$$
\Delta ^{\alpha \beta }\Delta _{\beta \mu }=\Delta _\mu ^\alpha ,~\Delta
^{\alpha \beta }U_\beta =0%
 \mbox{ and } \Delta _\mu ^\mu =n_{(a)}-1.
$$

With respect to a la--frame (\ref{dder}) we can introduce a particular
Lorentz system, called the locally anisotropic
 rest frame of the fluid when $%
U^\mu =\left( 1,0,0,...,0\right) $ and $\Delta _\mu ^\alpha =diag\left(
0,1,1,...,1\right) .$ So the pressure d--tensor was defined as to coincide
with the la--space part of the energy--momentum d--tensor with respect to a
locally anisotropic rest frame.

\subsection{Collisionless kinetic equation}

We start our proof by applying the identity
$$
\int ds\frac d{ds}\delta ^{n_{(a)}}\left( u^\alpha -u_{(l)}^\alpha
(s)\right) \delta ^{n_{(a)}}\left( p^\beta -p_{(l)}^\beta (s)\right) =0.
$$
Differentiating under this integral and taking into account the equation (%
\ref{maseq}) we get the relation
\begin{equation}
\label{aux1}\frac{\delta \left( p^\alpha \phi \right) }{\partial u^\alpha }%
+\frac \partial {\partial p_\beta }\left( ~^c\Gamma _{~\beta \varepsilon
}^\mu p_\mu p^\varepsilon \phi \right) =0.
\end{equation}
In our further considerations we neglect the interactions between the
particles and suppose that the metric of the background gravitational field $%
g_{\alpha \beta }$ does not depend on motion of particles. After averaging (%
\ref{aux1}) on paths we define the equation for the one--particle
distribution function
\begin{equation}
\label{1df}\frac{\delta \left( p^\alpha \Phi \right) }{\partial u^\alpha }%
+\frac \partial {\partial p_\beta }\left( ~^c\Gamma _{~\beta \varepsilon
}^\mu p_\mu p^\varepsilon \Phi \right) =0.
\end{equation}

As a matter of principle we can generalize the problem \cite{vlas} when
particles are considered to be also sources of gravitational field which is
self--consistently defined from the Einstein equations with the
energy--momentum tensor defined by the formula (\ref{fluxen}).

Taking into account the identity
$$
\frac{\delta p^\alpha }{\partial u^\alpha }+\frac \partial {\partial p_\beta
}(~^c\Gamma _{~\beta \varepsilon }^\mu p_\mu p^\varepsilon )=0
$$
we get from (\ref{1df}) the collisionless kinetic equation for the
distribution function $\Phi \left( u^\alpha ,p_\beta \right) $%
\begin{equation}
\label{clke}p^\beta ~\widehat{D}_\beta \Phi =0,
\end{equation}
where
\begin{equation}
\label{gcartan}\widehat{D}_\beta =\frac \delta {\partial u^\beta }-~^c\Gamma
_{~\beta \alpha }^\varepsilon p^\alpha \frac \partial {\partial
p^\varepsilon }
\end{equation}
generalizes the Cartan's covariant derivation \cite{cartan} for a space with
higher order an\-isot\-ro\-py (locally parametrized by supporting elements $%
\left( u^\alpha ,p_\beta \right) ),$ provided with a (extended to momenta
coordinates) higher order nonlinear connection
\begin{equation}
\label{hnc}\widehat{N}_\beta ^\varepsilon =\delta _a^\varepsilon \delta
_\beta ^iN_i^a-~^c\Gamma _{~\beta \alpha }^\varepsilon p^\alpha
\end{equation}
(on the geometry of higher order anisotropic spaces and superspaces and
possible applications in physics see \cite{ma,v1,v2}).

The equation (\ref{clke}) can be written in equivalent form for the
distribution function $f\left( u^\alpha ,p_{\widehat{\beta }}\right),$%
\begin{equation}
\label{clkeh}p^\lambda ~\widehat{D}_\lambda f=0,
\end{equation}
with the Cartan's operator defined on $n_{(a)}-1$ dimensional momentum space.

The kinetic equations (\ref{clke}) and, equivalently, (\ref{clkeh}) reflect
the conversation low of the quantity of particles in every volume of the
space of supporting elements which holds in absence of collisions.

Finally, in this subsection, we note that the kinetic equation (\ref{clke})
is formulated on the space of supporting elements $\left( u^\alpha ,p_\beta
\right) ,$ which is characterized by coordinate transforms
\begin{equation}
\label{ct}x^{i^{\prime }}=x^{i^{\prime }}\left( x^k\right) ,y^{a^{\prime
}}=y^aK_a^{a^{\prime }}\left( x^i\right)
\end{equation}
and
$$
p_{j^{\prime }}=p_i\frac{\partial x^i}{\partial x^{j^{\prime }}}%
,p_{a^{\prime }}=p_aK_{a^{\prime }}^a(x^i),
$$
where $K_a^{a^{\prime }}\left( x^i\right) $ and $K_{a^{\prime }}^a(x^i)$
take values correspondingly in the set of matrices para\-metr\-iz\-ing the
group of linear transforms $GL( m,{\R}^m ) ,$ where
$\R$ \ denotes the set of real numbers. In a particular case,
for dimensions $n=m$ we can parametrize
$$
K_i^{i^{\prime }}\left( x^i\right) =\frac{\partial x^{i^{\prime }}}{\partial
x^i} \mbox{ and } 
K_{i^{\prime }}^i(x^i)=\frac{\partial x^i}{\partial x^{i^{\prime }}}\mid
_{x^i=x^i(x^{i^{\prime }})}.
$$

A distinguished by higher order nonlinear connection (\ref{hnc}) tensor \\ $%
Q_{\beta _1\beta _2...\beta _q}^{\alpha _1\alpha _2...\alpha _r}
\left(u^\varepsilon ,p_\tau \right) $ being contravariant of rang $r$ and
covariant of rang $q$ satisfy the next transformation laws under chaingings
of coordinates of type (\ref{ct}):
\begin{equation}
\label{dtt}Q_{\beta _1^{\prime }\beta _2^{\prime }...\beta _q^{\prime
}}^{\alpha _1^{\prime }\alpha _2^{\prime }...\alpha _r^{\prime }}\left(
u^{\varepsilon ^{\prime }},p_{\tau ^{\prime }}\right) = \frac{\partial
u^{\alpha _1^{\prime }}}{\partial u^{\alpha _1}}...\frac{\partial u^{\alpha
_r^{\prime }}}{\partial u^{\alpha _r}}\frac{\partial u^\beta }{\partial
u^{\beta _1^{\prime }}}...\frac{\partial u^{\beta _q}}{\partial u^{\beta
_q^{\prime }}}Q_{\beta _1\beta _2...\beta _q}^{\alpha _1\alpha _2...\alpha
_r}\left( u^\varepsilon ,p_\tau \right) .
\end{equation}
Operators of type (\ref{hnc}) and transformations (\ref{ct}) and (\ref{dtt}%
), on first order of anisotropy, on tangent bundle $TM$ on a $n$ dimensional
manifold $M$ were considered by E. Cartan in his approach to Finsler
geometry \cite{cartan} and by A. A. Vlasov \cite{vlas} in order to formulate
the statistical kinetic theory on a Finsler geometry background. It should
be emphasized that the Cartan--Vlasov approach is to be applied even in
general relativity because the kinetic processes are to be examined in a
phase space provided with local coordinates $\left( u^\varepsilon ,p_\tau
\right) $. Our recent generalizations \cite{v1,v2} to higher order
anisotropy (including spinor and supersymmetric spaces) are to applied in
the case of models with nontrivial reductions (modelled by N--connections)
from higher dimensions to lower dimensional ones.

\section{Kinetic equation with pair collisions}

In this section we shall prove the relativistic kinetic equations for
one--particle distribution function $f(u^\alpha ,p_\beta )$ with pair
collisions \cite{chernikov}. We summarize the related results and generalize
the constructions for Minkowski spaces \cite{groot2}.

\subsection{Integral of collisions, differential cross--section and velocity
of transitions}

If collisions of particles are taken into account (for simplicity we shall
consider only pair collisions), the quantity of particles from a volume in
the space of supporting elements is not constant. We have to introduce a
source of particles $C\left( x,p\right) ,$ for instance, in the kinetic
equations (\ref{clkeh}),
\begin{equation}
\label{clkes}p^\alpha ~\widehat{D}_\alpha f=C\left( f\right) =C\left(
u,p\right) .
\end{equation}
The scalar function $C\left( u,p\right) $ is called the integral of
collisions (in the space of supporting elements). The value
$$
\triangle ^{n_{(a)}}u\frac{\triangle ^{n_{(a)}-1}p}{p^1}C\left( u,p\right)
$$
is the changing of the quantity of particles under pair collisions in a
region\\ $\triangle ^{n_{(a)}}u\triangle ^{n_{(a)}-1}p.$

Let us denote by
\begin{equation}
\label{colrate}\frac{W\left( p,p_{[1]}|p^{\prime },p_{[1]}^{\prime }\right)
}{p^1p_{[1]}^1p^{\prime 1}p_{[1]}^{\prime 1}}
\end{equation}
the probability of transition for two particles which before scattering have
the momenta $p_{\widehat{i}}$ and $p_{1\widehat{i}}$ and after scattering
the momenta $p_{\widehat{i}}^{\prime }$ and $p_{1\widehat{i}}^{\prime }$
with respective inaccuracies $\triangle ^{n_{(a)}-1}p^{\prime }$ and $%
\triangle ^{n_{(a)}-1}p_1^{\prime }.$ The function $W\left(
p,p_{[1]}|p^{\prime },p_{[1]}^{\prime }\right) $ (the so--called collision
rate) is symmetric on arguments $p,p_{[1]}$ and $p^{\prime },p_{[1]}^{\prime
}$ and describes the velocity of transitions with conservation of momenta of
type
$$
p^\alpha +p_{[1]}^\alpha =p^{\prime \alpha }+p_{[1]}^{\prime \alpha }
$$
when the conditions (\ref{maseq}) hold.

The number of binary collisions within a la--spacetime interval $\delta u$
around a point $u$ between particles with initial momenta in the ranges
$$
\left( \overrightarrow{p},\overrightarrow{p}+\delta \overrightarrow{p}%
\right) \mbox{ and } \left( \overrightarrow{p}_{[1]},\overrightarrow{p}%
_{[1]}+\delta \overrightarrow{p}_{[1]}\right)
$$
and final momenta in the ranges
$$
\left( \overrightarrow{p}^{\prime },\overrightarrow{p}^{\prime }+\delta
\overrightarrow{p}^{\prime }\right) \mbox{ and } \left( \overrightarrow{p}%
_{[1]}^{\prime }, \overrightarrow{p}_{[1]}^{\prime }+\delta \overrightarrow{p%
}_{[1]}^{\prime }\right)
$$
is given by
\begin{equation}
\label{colbin}f\left( u,p\right) f\left( u,p_{[1]}\right) W\left(
p,p_{[1]};p^{\prime },p_{[1]}^{\prime }\right) \delta \varsigma _p\delta
\varsigma _{p_{[1]}}\delta \varsigma _{p^{\prime }}\delta \varsigma
_{p_{[1]}^{\prime }}\delta u
\end{equation}
were, for instance, we abbreviated $\delta \varsigma _p=\delta ^{n_{(a)}}p/(%
\sqrt{|g|}p^{\overline{1}}).$

The collision integral of the Boltzmann equation in la--spacetime (\ref
{clkes}) is expressed in terms of the collision rate (\ref{colrate})
\begin{equation}
\label{colint}C\left( f\right) = \int f\left( u,p\right) f\left(
u,p_{[1]}\right) \ W\left( p,p_{[1]};p^{\prime },p_{[1]}^{\prime }\right)
\delta \varsigma _{p_{[1]}}\delta \varsigma _{p^{\prime }}\delta \varsigma
_{p_{[1]}^{\prime }}. \nonumber
\end{equation}
The collision integral is related to the differential cross section (see
below).

\subsection{The Cross Section in La--spacetime}

The number $n_{(bin)}$ of binary collisions per unit time and unit volume
when the initial momenta of the colliding particles lie in the ranges
$$
\left( \overrightarrow{p},\overrightarrow{p}+\delta \overrightarrow{p}%
\right) \mbox{ and } \left( \overrightarrow{p}_{[1]},\overrightarrow{p}%
_{[1]}+\delta \overrightarrow{p}_{[1]}\right)
$$
and the final momenta are in some interval $\varsigma $ in the space of
variables $\overrightarrow{p}^{\prime }$ and $\overrightarrow{p}%
_{[1]}^{\prime }$ is to be obtained from (\ref{colbin}) by dividing on $%
\delta u=\delta t\delta x^2...\delta x^ndy^1...dy^m$ and integrating with
respect to the primed variables,
\begin{equation}
\label{colbinvt}n_{(bin)} = f\left( u,p\right) f\left( u,p_{[1]}\right)
\times \delta \varsigma _p\delta \varsigma _{p_{[1]}}\int_\varsigma W\left(
p,p_{[1]};p^{\prime },p_{[1]}^{\prime }\right) \delta \varsigma _{p^{\prime
}}\delta \varsigma _{p_{[1]}^{\prime }}.
\end{equation}

Let us introduce an auxiliary velocity
\begin{equation}
\label{speed}v=F/(p^{\overline{1}}\ p_{[1]}^{\overline{1}})
\end{equation}
were the so--called M$\ddot o$ller flux factor is defined by
\begin{equation}
\label{moller}F=\left[ \left( p^\alpha p_{[1]\alpha }\right) ^2-\widetilde{m}%
^4\right] ^{1/2}.
\end{equation}
It may be verified that the speed $v$ reduces to the relative speed of
particles in a frame in which one of the particles initially is at rest. We
also consider the product of the number density of target particles with
momenta in the range $\left( \overrightarrow{p}_{[1]},\overrightarrow{p}%
_{[1]}+\delta \overrightarrow{p}_{[1]}\right) $ (given by $f\left(
u,p_{[1]}\right) \delta ^{n_{(a)}-1}p_{[1]})$ and the flux of incoming
particles with momenta in the range $\left( \overrightarrow{p},%
\overrightarrow{p}+\delta \overrightarrow{p}\right) $ (given by $f\left(
u,p\right) \delta ^{n_{(a)}-1}p\ v,$ where $v$ is the speed (\ref{speed}).
This product can be written
\begin{equation}
\label{product}Ff\left( u,p\right) f\left( u,p_{[1]}\right) \delta \varsigma
_p\delta \varsigma _{p_{[1]}}.
\end{equation}

By definition (like in usual the relativistic Boltzmann theory \cite
{chernikov,groot2,boisseau}) the cross section $\widetilde{\sigma }$ is the
division of the number (\ref{colbinvt}) to the number (\ref{product})
\begin{equation}
\label{cs}\widetilde{\sigma }=\frac 1F\int_\varsigma W\left(
p,p_{[1]};p^{\prime },p_{[1]}^{\prime }\right) \delta \varsigma _{p^{\prime
}}\delta \varsigma _{p_{[1]}^{\prime }}.
\end{equation}
The condition that collisions are local implies that the collision rate must
contain $n_{(a)}$ delta--functions
\begin{equation}
\label{deltafunction}W\left( p,p_{[1]};p^{\prime },p_{[1]}^{\prime }\right)
= w\left( p,p_{[1]};p^{\prime },p_{[1]}^{\prime }\right) \times
\delta^{n_{(a)}}\left( p+p_{[1]}-p^{\prime }-p_{[1]}^{\prime }\right) .
\end{equation}

Let ${\R}^{n_{(a)}-1}$ be a $(n_{(a)}-1)$--dimensional
Euclidean space enabled with an orthonormal basis $%
\left(e_1,e_2,...,e_{n_{(a)}-1}\right) $ and denote by $\left(
u^1,u^2,...,u^{n_{(a)}-1}\right) $ the Cartesian coordinates with respect to
this basis. For calculations on scattering of particles it is useful to
apply the system of spherical coordinates $(r_{n_{(a)}}=r,\theta
_{n_{(a)}-1},\theta _{n_{(a)}-2},...,\theta _1)$ associated with the
Cartesian coordinates $\left( u_1,u_2,...,\right).$ The volume element can
be expressed as%
$$
d^{n_{(a)}-1}u=(r)^{n_{(a)}-2}dr\ d^{n_{(a)}-2}\Omega _\theta ,
$$
where the element of solid angle, the $(n_{(a)}-1)$--spherical element, is
given by
\begin{eqnarray}
&d^{n+m-2}&\Omega _\theta  = \sin ^{n+m-3}\theta _{n+m-2}\cdot \sin
^{n+m-4}\theta _{n+m-3}\cdot ... \nonumber \\
 &  & \sin ^2\theta _3\cdot \sin \theta_2  \times
d\theta _{n+m-2}\cdot ... \cdot d\theta _2\cdot d\theta _1.
 \label{sa}
\end{eqnarray}
In our further considerations we shall consider that the Cartesian and
spherical $(n_{(a)}-1)$--dimensional coordinates are given with respect to a
la--frame of type (\ref{dder}).

In order to eliminate the delta functions from (\ref{deltafunction}) put
into (\ref{cs}) we fix as a reference frame the center of mass frame for the
collision between two particles with initial momenta $\overrightarrow{p}$
and $\overrightarrow{p}_{[1]}.$ The quantities defined with respect to the
center of mass frame will be enabled with the subindex $CM.$ One denotes by $%
\overrightarrow{p}_{CM}$ the polar axis and characterizes the directions $%
\overrightarrow{p}_{CM}^{\prime }$ of the outgoing particles with respect to
polar axis by means of generalized spherical coordinates. In this case
\begin{equation}
\label{mecm}\delta ^{n_{(a)}-1}p_{CM}^{\prime }=\left| \overrightarrow{p}%
_{CM}^{\prime }\right| ^{n_{(a)}-2}~d\left| \overrightarrow{p}_{CM}^{\prime
}\right| d\Omega _{CM}
\end{equation}
where $d\Omega _{CM}$ is given by the formula (\ref{sa}).

The total $\left( n_{(a)}+1\right) $--momenta before and after collision in
la--spacetime are given by d--vectors
\begin{equation}
\label{totmom}P^\alpha = p^\alpha +p_{[1]}^\alpha, \qquad P^{\prime \alpha }
= p^{\prime \alpha }+p_{[1]}^{\prime \alpha }.
\end{equation}
Following from
$$
P_{CM}^{\prime 1}=2p_{CM}^{\prime 1}=2\sqrt{\left| \overrightarrow{p}%
_{CM}^{\prime }\right| ^2+\widetilde{m}^2}
$$
we have
$$
\left| \overrightarrow{p}_{CM}^{\prime }\right| ^2~d\left| \overrightarrow{p}%
_{CM}^{\prime }\right| =\frac 14P_{CM}^{\prime 1}~dP_{CM}^{\prime 1}.
$$
In result the formula for the volume element (\ref{mecm}) transforms into%
$$
\delta ^{n_{(a)}-1}p_{CM}^{\prime }=\frac 14\left| \overrightarrow{p}%
_{CM}^{\prime }\right| ^{n_{(a)}-2}~P_{CM}^{\prime 1}~dP_{CM}^{\prime
1}~d\Omega _{CM}.
$$

Inserting this collision rate (\ref{deltafunction}) into (\ref{cs}) we
obtain an integral which can be rewritten (see the Appendices 1 and 7 to the
paper \cite{boisseau} for details on transition to the center of mass
variables; in la--spacetimes one holds good similar considerations with that
difference that me must work with respect to la--frames (\ref{dder}) and (%
\ref{ddif}) and d--metric (\ref{dmetric}) )
\begin{equation}
\label{cscm}\sigma =\frac{F^{n_{(a)}-4}}{E^{n_{(a)}-2}}\int_\varsigma w\
d\Omega _{CM}
\end{equation}
where $E=\sqrt{P^\alpha P_\alpha },,$ is the total energy (devided by $c=1).$

The differential section in the center of mass system is found from (\ref
{cscm})
$$
\frac{\delta \sigma }{d\Omega }\mid _{_{CM}}=\frac{F^{n_{(a)}-4}}{%
E^{n_{(a)}-2}}w
$$
which allow to express the collision rate (\ref{deltafunction}) as
\begin{equation}
\label{deltafunctionds}W \left( p,p_{[1]};p^{\prime },p_{[1]}^{\prime
}\right) = \frac{E^{n_{(a)}-2}}{F^{n_{(a)}-4}} \times \frac{\delta \sigma }{%
d\Omega }\mid _{_{CM}}\cdot \delta ^{n_{(a)}}\left( p+p_{[1]}-p^{\prime
}-p_{[1]}^{\prime }\right) .
\end{equation}
We shall use the formula (\ref{deltafunctionds}) in order to compute the
transport coefficients.

\section{Equilibrium States in La--Spacetimes}

When the system is in equilibrium we can derive an expression for the
particle distribution function $f\left( u,p\right) =f_{(eq)}\left(
u,p\right) $ in a similar way as for locally isotropic spaces and write
\begin{equation}
\label{distrfuncteq}f_{(eq)}\left( u,p\right) =\frac 1{(2\pi \hbar
)^{n_{(a)}-1}}\exp \left( \frac{\mu -p^iU_i-p^aU_a}{k_BT}\right) ,
\end{equation}
where $k_B$ is Boltzmann's constant and $\hbar $ is Planck's constant
divided by $2\pi $ and $\mu =\mu \left( u\right) $ and $T=T\left( u\right) $
are respectively the thermodynamic potential and temperature. For
simplicity, in this section we shall omit explicit dependencies of $\mu $
and $T$ on la--spacetime coordinates $u$. Our thermodynamic systems will be
considered in local equilibrium in a vicinity of a point $u_{[0]}$ with
respect to a rest frame locally adapted to N--connection structure, like (%
\ref{dder}) and (\ref{ddif}). In the simplest case the $n+m$ splitting is
trivially given by a N--connection with vanishing curvature (\ref{ncurv}).
For such conditions of trivial la--spacetime $\mu $ and $T$ can be
considered as constant values.

\subsection{Particle Density}

The formula relating the particle density $\widetilde{n},$ temperature $T,$
and thermodynamic potential $\mu $ is obtained by inserting (\ref
{distrfuncteq}) into (\ref{pdens}), with (\ref{fluxpart}). Choosing the
locally adapted to the N--connection (\ref{dder}) to be the rest frame, when
$U_\mu =\left( 1,0,\ldots ,0\right) $ the calculus is to be performed as for
isotropic $n_{(a)}$ dimensional spaces \cite{boisseau}. The integral for the
number of particles per unit volume in the rest la--frame is
\begin{equation}
\label{parnumvol}\widetilde{n}\left( \mu ,T\right) = \frac 1{\left( 2\pi
\hbar \right) ^{n_{(a)}-1}}\exp \left( \frac \mu {k_BT}\right) \times \int
p^{\overline{1}} \exp \left( -\frac{p^{\overline{1}}}{k_BT}\right) \frac{%
d^{n_{(a)}-1}p}{p^{\overline{1}}},
\end{equation}
with
\begin{equation}
\label{energy}p^{\overline{1}}=\left( \left| \overrightarrow{p}\right| ^2+%
\widetilde{m}^2\right) ^{1/2}
\end{equation}
and depends only on length of $n_{(a)}-1$ dimensional d--vector $%
\overrightarrow{p}.$ By applying spherical coordinates, when
$$
d^{n_{(a)}-1}p=\left| \overrightarrow{p}\right| ^{n_{(a)}-2}~d\left|
\overrightarrow{p}\right| ~d\Omega
$$
and $d\Omega $ is given by the expression (\ref{sa}), and differentiating on
radius $\rho $ the well known formula for the volume $V_{n_{(a)}-1}$ in a $%
n_{(a)}-1$ dimensional Euclidean space
\begin{equation}
\label{sphvol}V_{n_{(a)}-1}=\frac{\left( n_{(a)}-1\right) \pi ^{\left(
n_{(a)}-1\right) /2}}{\Gamma [(n_{(a)}+1)/2]}\rho ,
\end{equation}
where $\Gamma $ is the Euler gamma function, then putting $\rho =1$ we get
\begin{equation}
\label{sphint}\int d\Omega =\frac{\left( n_{(a)}-1\right) \pi ^{\left(
n_{(a)}-1\right) /2}}{\Gamma [(n_{(a)}+1)/2]}.
\end{equation}
We note that from (\ref{energy}) one follows
$$
\left| \overrightarrow{p}\right| ~d\left| \overrightarrow{p}\right| =p^{%
\overline{1}}~dp^{\overline{1}}
$$
and we can transform (\ref{parnumvol}) into an integral with respect to the
angular variables through the replacement
$$
\frac{d^{n_{(a)}-1}p}{p^{\overline{1}}}\rightarrow \frac{\left(
n_{(a)}-1\right) \pi ^{\left( n_{(a)}-1\right) /2}}{\Gamma [(n_{(a)}+1)/2]}%
\left| \overrightarrow{p}\right| ^{n_{(a)}-3}dp^{\overline{1}}.
$$
Introducing the dimensionless quantities
\begin{equation}
\label{dimlessvar}\xi = p^{\overline{1}}/k_BT, \mbox{ and } \vartheta =
\widetilde{m}/k_BT,
\end{equation}
for which, respectively,
\begin{equation}
\label{aux01}dp^{\overline{1}} = k_BT~d\xi, \mbox{ and } \left|
\overrightarrow{p}\right| = k_BT\sqrt{\xi ^2-\vartheta ^2},
\end{equation}
the integral (\ref{parnumvol}) is computed
\begin{eqnarray} \label{nmut}
{\widetilde n}\left( \mu ,T\right) & = & 2^{n_{(a)}/2}
\left( \frac{\pi \widetilde{m}}{2\pi \hbar }
\right) ^{n_{(a)}-1}\left( \frac{k_BT}{\widetilde{m}}\right) ^{\left(
 n_{(a)}-2\right) /2} \\
 { } & \times & \exp \left( \frac \mu {k_BT}\right) K_{(n_{(a)}+1)/2}\left(
\frac{\widetilde{m}}{k_BT}\right) , \nonumber
\end{eqnarray}
where the modified Bessel function of the second kind of order $n_{(a)}/2$
has the integral representation
\begin{equation}
\label{bessint}K_s\left( \eta \right) =\frac{\sqrt{\pi }}{\left( 2\eta
\right) ^s\Gamma \left( s+\frac 12\right) }\int_\eta ^\infty e^{-q}\left(
q^2-\eta ^2\right) ^{s-1/2}dq.
\end{equation}

We note the dependence of (\ref{nmut}) (but not of (\ref{aux01})) on $m$
anisotropic parameters (coordinates). The formulas proved in this subsection
transforms into locally anisotropic ones \cite{boisseau} if the
N--connection is fixed to be trivial (with vanishing N--connection curvature
(\ref{ncurv})) and the d--metric (\ref{dmetric}) transforms into a usual
(pseudo) Riemannian one.

\subsection{Average energy and pressure}

The average energy per particle, for a system in equilibrium, can be
calculated by introducing the distribution function (\ref{distrfuncteq})
into (\ref{edensp}) and applying the formulas (\ref{fluxen}), (\ref{edens})
and (\ref{nmut}). In terms of dimensionless variables (\ref{dimlessvar}) we
have
\begin{eqnarray}
\varepsilon \left(\mu ,T\right) & = & \frac{\left( k_BT\right)
^{n_{(a)}}}{\left( 2\pi \hbar \right) ^{n_{(a)}-1}}\
  \frac{\left(
n_{(a)}-1\right) \pi ^{\left( n_{(a)}-1\right) /2}}{\Gamma [(n_{(a)}+1)/2]}
 \nonumber \\
 { } & \times &
\exp \left( \frac \mu {k_BT}\right) \int_\vartheta ^\infty \left( \xi
^2-\vartheta ^2\right) ^{(n_{(a)}-3)/2}\xi ^2d\xi . \nonumber
\end{eqnarray}
After carrying out partial integrations together with applications of the
formula%
$$
\left( q^2-\zeta ^2\right) ^{\left( s-2\right) /2}=\frac 1s\frac d{dq}\left[
\left( q^2-\zeta ^2\right) ^{s/2}\right]
$$
we obtain
\begin{eqnarray}
\varepsilon \left( \vartheta \right) & = &
\pi ^{\left( n_{(a)}-2\right) /2}\frac{\widetilde{m}^{n_{(a)}}}
{\left( 2\pi \hbar \right) ^{n_{(a)}-1}}\left( \frac
2\vartheta \right) ^{n_{(a)}/2} \label{edensbess} \\
 { } & \times &
\exp \left( \frac \mu {k_BT}\right) \left[
\vartheta K_{(n_{(a)}+2)/2}\left( \vartheta \right) -K_{n_{(a)}/2}\left(
\vartheta \right) \right] .\nonumber
\end{eqnarray}

Dividing the energy density (\ref{edensbess}) to the particle density (\ref
{parnumvol}) and substituting the inverse dimensionless temperature $%
\vartheta =\vartheta \left( T\right) $ (\ref{dimlessvar}) we get the energy
per particle (\ref{edensp})
\begin{equation}
\label{trmeqst}e\left( T\right) =k_BT\left( \frac{\widetilde{m}}{k_BT}\frac{%
K_{(n_{(a)}+2)/2}\left( \frac{\widetilde{m}}{k_BT}\right) }{%
K_{n_{(a)}/2}\left( \frac{\widetilde{m}}{k_BT}\right) }-1\right) .
\end{equation}
This formula is the thermal equation of state of a equilibrium system having
$m$ anisot\-ro\-py parameters (we remember that $n_{(a)}=n+m)$ with respect
to a la--frame (\ref{dder}).

The pressure d--tensor is to be computed by substituting (\ref{distrfuncteq}%
) into (\ref{pressproj}), applying also the integral (\ref{fluxen}). We get
\begin{equation}
\label{press}P^{\alpha \beta }=-k_BT~\widetilde{n}\left( \mu ,T\right)
\Delta ^{\alpha \beta },
\end{equation}
were, by definition, the coefficients before $\Delta ^{\alpha \beta }$
determine the pressure
\begin{equation}
\label{pressure}\widetilde{P}=k_BT~\widetilde{n}\left( \mu ,T\right) .
\end{equation}

So, for a system of $n_{(a)}-1=n+m-1$ dimensions the expression (\ref
{pressure}) defines the equation of state of ideal gas of particles with
respect to a la--frame, having $m$ anisotropic parameters.

\subsection{Enthalpy, specific heats and entropy}

The average enthalpy per particle
\begin{equation}
\label{enthalpyp}h\left( T\right) =e\left( T\right) +\widetilde{P}\left( \mu
,T\right) /\widetilde{n}\left( \mu ,T\right)
\end{equation}
is computed directly by substituting in this formula the values (\ref
{trmeqst}),(\ref{pressure}) and (\ref{nmut}). The result is
\begin{equation}
\label{enthb}h\left( T\right) =\widetilde{m}\frac{K_{(n_{(a)}+2)/2}\left(
\frac{\widetilde{m}}{k_BT}\right) }{K_{n_{(a)}/2}\left( \frac{\widetilde{m}}{%
k_BT}\right) }.
\end{equation}

Using (\ref{trmeqst}) and (\ref{enthb}) we can compute respectively the
specific heats at constant pressure and at constant volume%
$$
c_p=\left( \frac{\partial h}{\partial T}\right) _p\mbox{ and } c_v=\left(%
\frac{\partial e}{\partial T}\right) _v.
$$

Subtracting of $h$ and $e,$ after carrying out the differentiation with
respect to $T,$ one finds Mayer's relation,%
$$
c_p-c_v=k_B
$$
and, introducing the adiabatic constant $\gamma =c_p/c_v,$
\begin{equation}
\label{adconst}\frac \gamma {\gamma -1}=\vartheta ^2+\left( n_{(a)}+1\right)
\frac h{k_BT}-\left( \frac h{k_BT}\right) ^2.
\end{equation}
The relation (\ref{adconst}) can be proven by straightforward calculations
by using the properties of Bessel's function (\ref{bessint}) (see a similar
proof for isotropic spacetimes in \cite{boisseau}).

The entropy per particle is introduced as in the isotropic case
\begin{equation}
\label{entropy}s=\frac{h-\mu }T.
\end{equation}
We have to insert the values of $\mu $ (found from (\ref{nmut})
\begin{equation}
\label{mupot}\mu \left( \widetilde{n},T\right) = k_BT \times \ln \frac{%
\left( 2\pi \hbar \right) ^{n_{(a)}-1}\widetilde{n}}{(2\widetilde{m}%
)^{n_{(a)}/2}(\pi k_BT)^{(n_{(a)}-2)/2}K_{n_{(a)}/2}\left( \frac{\widetilde{m%
}}{k_BT}\right) }
\end{equation}
and of $h$ (see (\ref{enthb}) in order to obtain the dependence $s=s\left(
T\right) \,$ (for simplicity, we omit this cumbersome formula).

\subsection{Low and high energy limits}

Let us investigate the non--relativistic local\-ly an\-isot\-rop\-ic limit
by applying the asymptotic formula \cite{abram}
$$
K_s\left( \vartheta \right) \sim \sqrt{\frac \pi {2\vartheta }}e^{-\vartheta
}
$$
valid for large $\vartheta .$ The formula (\ref{trmeqst}) gives%
$$
e\sim \widetilde{m}+\frac{n+m-1}2k_BT.
$$
Thus, each isotropic and anisotropic dimension contributes with $k_BT/2$ to
the average energy per particle.

Now, we consider high temperatures. For $\vartheta \rightarrow 0$ one holds
the asymptotic formula \cite{abram}%
$$
K_s\left( \vartheta \right) \sim 2^{s-1}\Gamma \left( s\right) \vartheta
^{-s}
$$
and the particle number density (\ref{nmut}), the energy (\ref{trmeqst}),
enthalpy (\ref{enthb}) and entropy (\ref{entropy}) can be approximated
respectively (we state explicitly the dimensions $n+m)$ by
\begin{eqnarray} \label{ultrens}
{\widetilde n}\left( \mu ,T\right) & = &
2^{n+m-1}\pi ^{n+m-2}\Gamma \left( \frac{n+m}2\right)
 \times \left( \frac{k_BT}{2\pi \hbar }\right) ^{n+m-1}
  \exp \left( \frac \mu{k_BT}\right) , \\
e & = & \left( n+m-1\right) k_BT,~h=\left( n+m\right) k_BT, \nonumber \\
s & = & \left(n+m\right) k_B. \nonumber
\end{eqnarray}
In consequence, the corresponding specific heats and adiabatic constant are
$$
c_p=\left( n+m\right) k_B,c_v=\left( n+m-1\right) k_B \mbox{ and } \gamma =
\frac{n+m}{n+m-1}.%
$$
These formulas imply that for large $ T$ the spacetime anisotropy (very
possible at the beginning of our Universe) could modify substantially the
thermodynamic parameters.

Finally we emphasize that in the ultrarelativistic limit
\begin{equation}
\label{ultrentr}s = 2^{n+m-1}\pi ^{n+m-2}\left( n+m\right) \times \Gamma
\left( \frac{n+m}2\right) \left( \frac{k_BT}{2\pi \hbar }\right) ^{n+m-1}%
\frac{k_B}{\widetilde{n}}.
\end{equation}
We also note that our treatment was based on Maxwell--Boltzmann instead of
Bose--Einstein statistics.

\section{Linearized Locally Anisotropic \protect\newline Transport Theory}

For systems with local anisotropy outside equi\-lib\-ri\-um the
dis\-tri\-bu\-ti\-on function\ $f_{(eq)}(u,p)$ (\ref{distrfuncteq}) must be
generalized to another one, $f(u,p),$ solving the kinetic la--equation (\ref
{clkes}). We follow a standard procedure of linearization and construction
of solutions of kinetic equations by generalizing to la--spacetimes the
Chapman--Enskog approach (we shall extend to locally anisotropic backgrounds
the results presented in \cite{groot2}).

We write
\begin{equation}
\label{dfche}f\left( u,p\right) =f_{[0]}\left( u,p\right) \left[ 1+\varphi
\left( u,p\right) \right]
\end{equation}
with the lowest order of approximation to $f$ taken similarly to (\ref
{distrfuncteq})
\begin{equation}
\label{fzero}f_{[0]}\left( u,p\right) = \frac 1{(2\pi \hbar )^{n_{(a)}-1}}
\times \exp \left( \frac{\mu (u)-p^iU_i(u)-p^aU_a(u)}{k_BT(u)}\right)
\end{equation}
where the constant variables of a trivial equilibrium $\mu ,T$ and $U_\alpha
\,$ are changed by some local conterparts $\mu \left( u\right) ,T\left(
u\right) $ and $U_\alpha \left( u\right) .$ Outside equilibrium the first
two dependencies $\mu \left( u\right) $ and $T\left( u\right) $ are defined
respectively from relations (\ref{mupot}) and (\ref{trmeqst}).

\subsection{Linearized Transport Equations}

The so--called deviation function $\varphi \left( u,p\right) $ from (\ref
{dfche}) describes the deformation by nonequilibrium flows of $%
f_{[0]}\left(u,p\right) $ into $f\left( u,p\right) .$ It is considered that
in equilibrium $\varphi $ vanishes and not too far from equilibrium states
it must be small. The Chapman--Enskog method states that after substituting $%
f_{[0]}$ into the left side and $f_{[0]}\left( 1+\varphi \right) $ into the
right hand of (\ref{clkes}) we shall neglect the quadratic and higher terms
in $\varphi .$ In result we obtain a linearized equation for the deviation
function (for simplicity, hereafter we shall not point to the explicit
dependence of functions, kinetic and thermodynamic values on spacetime
coordinates)%
$$
-p^\alpha \widehat{D}_\alpha f_{[0]}=f_{[0]}L[\varphi ]
$$
where, having introduced (\ref{fzero}) into (\ref{colint}), we write for the
right side
$$
L[\varphi ]=-p^\alpha \widehat{D}_\alpha \left( \frac{\mu -p^iU_i-p^aU_a}{%
k_BT}\right) .
$$
The left side is to be computed by applying the generalized Cartan
derivative (\ref{gcartan}).

We introduce a locally anisotropic generalization of the gradient operator
by considering the operator (in brief, la--gradient)
\begin{equation}
\label{lagradient}\nabla _\alpha =\triangle _\alpha ^\beta ~D_\beta
\end{equation}
with $\triangle _\alpha ^\beta $ given by (\ref{proj}) (we use the a
d--covariant derivative defined by a d--connection (\ref{dcon}) instead of
partial and/or isotropic derivatives for isotropic spaces \cite
{groot2,boisseau}). In a local rest frame of the isotropic fluid $\nabla
_\alpha \rightarrow \left( 0,\partial _1,\partial _2,...\right) ,$ i.e. in
the flat isotropic space the la--gradient reduces to the ordinary space
gradient. By applying $\nabla _\alpha $ we can eliminate the time
la--derivatives%
\begin{eqnarray} \label{transpart}
k_BTL[\varphi ] & = &
p^\alpha p^\beta \nabla _\alpha U_\beta -Tp^\alpha \nabla
_\alpha \left( \frac \mu T\right)   -p^\alpha p^\beta U_\alpha
\left( \frac{\nabla _\beta T}T-\frac{\nabla _\beta p}{h\widetilde{n}}\right)
   \\
{} &  & + [ \left( \gamma -1\right) p^\alpha U_\alpha +T^2\left( \gamma
-1\right) \frac \partial {\partial T}\left( \frac \mu T\right)
  + {\widetilde n}\left(
\frac{\partial \mu }{\partial n}\right) ] p^\varepsilon U_\varepsilon
\nabla ^\upsilon U_\upsilon  . \nonumber
\end{eqnarray}

We can verify, using the expression $\mu =\mu \left( \widetilde{n},T\right) $
from (\ref{mupot}) that one holds the equalities
$$
\frac \partial {\partial T}\left( \frac \mu T\right) =-\frac e{T^2},~\quad
\frac{\partial \mu }{\partial \widetilde{n}}=\frac{k_BT}{\widetilde{n}}
$$
and
$$
\frac 1{\widetilde{n}}\nabla _\alpha ~p=h\frac{\nabla _\alpha ~T}T+T\nabla
_\alpha \left( \frac \mu T\right) .
$$
Introducing these expressions into (\ref{transpart}) we get%
\begin{eqnarray}
&k_BT& L[\varphi ]  =  \Xi X+
\left( h-p^\alpha U_\alpha \right) \triangle
_\varepsilon ^\beta p_\beta X^\beta  \nonumber \\
 & & +  \left( \triangle _\varepsilon ^\beta
\triangle _\gamma ^\alpha -\frac 1{n+m-1}\triangle _{\varepsilon \gamma
}\triangle ^{\beta \alpha }\right) p_\beta p_\alpha ~^{(0)}X^{\varepsilon
\gamma }, \nonumber
\end{eqnarray}
where
\begin{equation}
\label{force1}\Xi = \left( \frac{n+m}{n+m-1}-\gamma \right) \left( p^\alpha
U_\alpha \right) ^2 +\left[ \left( \gamma -1\right) h-\gamma k_BT\right]
p^\alpha U_\alpha -\frac{\widetilde{m}^2}{n+m-1}
\end{equation}
and there were considered forces deriving the system towards equilibrium
$$
X=-\nabla ^\mu U_\mu ,~X^\alpha =\frac{\nabla ^\alpha ~T}T-\frac{\nabla
^\alpha ~p}{\widetilde{n}h}
$$
and%
$$
^{(0)}X^{\varepsilon \gamma }=\frac 12\left( \nabla ^\varepsilon U^\gamma
+\nabla ^\gamma U^\varepsilon \right) -\frac 1{n+m-1}\triangle ^{\varepsilon
\gamma }\nabla ^\mu U_\mu .
$$

We note that the values $k_BT~L[\varphi ],\Xi $ and $^{(0)}X^{\varepsilon
\gamma }$ depend explicitly on the dimensions of the base subspace, $n,$ and
of the fiber subspace, $m.$ The anisotropy also modify both the
thermodynamic and kinetic values via operators $\nabla ^\mu $ and $\triangle
_\varepsilon ^\beta $ which depend explicitly on d--metric and d--connection
coefficients.

\subsection{On the solution of locally anisotropic transport equations}

Let us suppose that is known the solution of these three equations:
\begin{eqnarray}
\label{transpeq}k_BTL[A\left( p\right) ] &= &\Xi , \\
 k_BTL[B\left( p\right)
\triangle _\alpha ^\beta ~p_\beta ] & = &\left( h-p^\alpha U_\alpha \right)
\triangle _\alpha ^\beta ~p_\beta \nonumber
\end{eqnarray}
and%
$$
k_BTL\left[ C\left( p\right) \left( \triangle _\alpha ^\beta \triangle
_\varepsilon ^\sigma -\frac 1{n+m-1}\triangle _{\alpha \varepsilon
}\triangle ^{\beta \sigma }\right) ~p_\beta ~p_\sigma \right]
$$
$$
=\left( \triangle _\alpha ^\beta \triangle _\varepsilon ^\sigma -\frac
1{n+m-1}\triangle _{\alpha \varepsilon }\triangle ^{\beta \sigma }\right)
~p_\beta ~p_\sigma
$$
for some functions $A\left( p\right) ,B\left( p\right) $ and $C\left(
p\right) .$ Because the integral operator $L$ is linear the linear
combination of thermodynamic forces
$$
\varphi =AX+B_\alpha X^\alpha +C_{\alpha \beta }~^{(0)}X^{\alpha \beta },
$$
where
$$
B_\alpha =B\left( p\right) \triangle _\alpha ^\beta ~p_\beta
$$
and
$$
C_{\alpha \beta }=B\left( p\right) \left( \triangle _\alpha ^\varepsilon
\triangle _\beta ^\sigma -\frac 1{n+m-1}\triangle _{\alpha \beta }\triangle
^{\varepsilon \sigma }\right) ~p_\varepsilon ~p_\sigma ,
$$
is a solution for the deviation function $\varphi .$

It can also be verified that every function of the form $\varphi =a+b_\alpha
p^\alpha $ with some parameters $a$ and $b_\alpha $ not depending on $%
p^\alpha $ is a solution of the homogeneous equation $L\left[ \varphi
\right] =0.$ In consequence, it was proved (see \cite{dijkstra} and \cite
{boisseau}) that the scalar parts of $A\left( p\right) $ and $B\left(
p\right) $ are determined respectively up to functions of the forms $\varphi
=a+b_\alpha p^\alpha $ and $b^\alpha p_\alpha .$ We emphasize that solutions
of the transport equations (\ref{transpeq}) (see next subsections) will be
expressed in terms of $A,B,$ and $C.$

\subsection{Linear laws for locally anisotropic non--equilibrium \protect
\newline thermodynamics}

Let $U_\beta \left( u\right) $ be a velocity field of Landau--Lifshitz type
characterizing a locally an\-isot\-rop\-ic fluid flow. The heat flow and the
viscous pressure d--tensor with respect to a such la--field are
correspondingly defined similarly to \cite{groot2} but in terms of objects
on la--spacetime (see formulas (\ref{enthb}), (\ref{press}), (\ref{proj}), (%
\ref{pressproj}), (\ref{fluxen}), (\ref{fluxpart}), (\ref{dcon}), (\ref{dder}%
), (\ref{ddif}), (\ref{dmetric})
\begin{eqnarray}
\label{nethp}
I_{(heat)}^\alpha & = &\left( U_\alpha \Upsilon ^{\alpha \beta
}-hn^\beta \right) \triangle _\beta ^\varepsilon , \\
\Pi ^{\alpha \beta } & = & P^{\alpha \beta }+p\triangle ^{\alpha \beta }.
 \nonumber
\end{eqnarray}
It should be noted that if the Landau--Lifshitz condition (\ref{llcond}) is
satisfied the first term in $I_{(heat)}^\nu $ vanishes and the heat flow is
the enthalpy carried away by the particles. The pressure $P\left( u\right) $
is defined by $\widetilde{n}\left( u\right) k_BT\left( u\right) ,$ see (\ref
{pressure}). Inserting (\ref{dfche}) and (\ref{fzero}) into (\ref{nethp}) we
prove (see locally isotropic cases in \cite{groot2,boisseau}) the linear
laws of locally anisotropic non--equilibrium thermodynamics%
$$
I_{(heat)}^\nu =\lambda TX^\nu
$$
and
$$
\Pi ^{\alpha \beta }=2\eta ~^{(0)}X^{\alpha \beta }-\eta _{(v)}X\triangle
^{\varepsilon \sigma }
$$
with the transport coefficients (the heat conductivity $\lambda ,$ the shear
viscosity $\eta ,$ and the volume viscosity coefficient $\eta _{(v)};$\ in
order to compare with formulas from \cite{boisseau} we shall introduce
explicitly the light velocity constant $c)$ defined as
\begin{eqnarray}
\lambda & = &-\frac c{(n+m-1)T}\int p^\sigma B_\varepsilon \triangle _\sigma
^\varepsilon \left( h-p^\nu U_\nu \right) f_{[0]}d\varsigma _p,%
\nonumber \\
\eta _{(v)} & = & -\frac c{n+m-1}\int p_\varepsilon p_\sigma \triangle
^{\varepsilon \sigma }Af_{[0]}d\varsigma _p,  \label{transpcoef1} \\
\eta & = & \frac c{(n+m)\left( n+m-4\right) }
\int p^\sigma p^\varepsilon C_{\nu
\mu }\triangle _{\sigma \varepsilon }^{\nu \mu }f_{[0]}d\varsigma _p,
\nonumber
\end{eqnarray}
where one uses the d--tensor%
$$
\triangle _{\sigma \varepsilon }^{\nu \mu }=\frac 12\left( \triangle _\sigma
^v\triangle _\varepsilon ^\mu -\triangle _\varepsilon ^v\triangle _\sigma
^\mu \right) -\frac 1{n+m-1}\triangle ^{\nu \mu }\triangle _{\sigma
\varepsilon }
$$
with the properties
$$
\triangle _{\rho \sigma }^{\nu \mu }\triangle _{\phi \varepsilon }^{\rho
\sigma }=\triangle _{\phi \varepsilon }^{\nu \mu }
$$
and
$$
\triangle_{\rho \sigma }^{\rho \sigma }=-1+\frac{(n+m)(n+m-1)}2.
$$

The main purpose of non--equilibrium thermodynamics is the calculation of
transport coefficients. With respect to la--frames the formulas are quite
similar with those for isotropic spaces with that difference that we have to
consider the values as d--tensors and take into account the number $m$ of
anisotropic variables.

On both type of locally isotropic and anisotropic spacetimes one holds the
so--called conditions of fit (see, for instance, \cite{boisseau})%
$$
\int p_\varepsilon U^\varepsilon Af_{[0]}d\varsigma _p=0 \mbox{ and } \int
\left( p_\varepsilon U^\varepsilon \right) ^2Af_{[0]}d\varsigma _p=0
$$
which allow us to write the volume viscosity (see (\ref{force1}))%
$$
\eta _{(v)}=c\int \Xi Af_{[0]}d\varsigma _p.
$$

For some d--tensors $H^{\alpha _1,\ldots \alpha _q}\left(p\right) $ and $%
S_{\alpha _1,\ldots \alpha _q}(p)$ we define the symmetric bracket
$$
\left\{ H,S\right\} =\frac 1{\widetilde{n}^2}\int H^{\alpha _1,\ldots \alpha
_q}\left( p\right) L\left[ S_{\alpha _1,\ldots \alpha _q}(p)\right]
f_{[0]}d\varsigma _p
$$
where $\widetilde{n}$ is the particle density and $L$ is a linearized
operator. In terms of such brackets the coefficients (\ref{transpcoef1}) can
be rewritten in an equivalent form%
\begin{eqnarray}
\lambda & = & -\frac{ck_B\widetilde{n}^2}{n+m-1}\left\{ B^\alpha ,B_\alpha
\right\} ,\label{transpcoef2} \\
\eta & = & \frac{ck_BT\widetilde{n}^2}{(n+m-1)^2-3}\left\{ C^{\alpha \beta
},C_{\alpha \beta }\right\} , \nonumber \\
\eta _{(v)} & = & ck_BT\widetilde{n}^2\left\{ A,A\right\} . \nonumber
\end{eqnarray}

The addition of invariants of type $\varphi =a+b_\alpha p^\alpha $ to some
solutions for $A\left( p\right) ,$ $B_\alpha \left( p\right) ,$ or $%
C_{\alpha \beta }\left( p\right) $ does not change the values of the
transport coefficients.

\subsection{Integral and algebraic equations}

The solutions of transport equations (\ref{transpeq}) are approximated \cite
{boisseau} by considering power series on $\zeta =p^\alpha U_\alpha /k_BT$
for functions
$$
A\left( \zeta \right) =\sum\limits_{\widehat{a}=2}^\infty A_{\widehat{a}%
}\zeta ^{\widehat{a}},B\left( \zeta \right) =\sum\limits_{\widehat{b}%
=1}^\infty B_{\widehat{b}}\zeta ^{\widehat{b}},C\left( \zeta \right)
=\sum\limits_{\widehat{c}=0}^\infty C_{\widehat{c}}\zeta ^{\widehat{c}}.
$$
The starting values $\widehat{a}=2$ and $\widehat{b}=1$ have been introduced
with the aim to determine the scalar functions $A\left( \zeta \right) $ and $%
B\left( \zeta \right) ,$ respectively, up to contributions of the forms $%
a+b_\alpha p^\alpha $ and $b^\alpha p_\alpha .$ Inserting these power series
into the integral equations (\ref{transpeq}), multiplying respectively on $%
\zeta ^{\widehat{s}}f_{[0]},\zeta ^{\widehat{s}}p^\alpha f_{[0]},\,$ and $%
\zeta ^{\widehat{s}}p^\alpha p^\beta f_{[0]},$ after integrating on $%
d\varsigma _p$ we get%
$$
\sum\limits_{\widehat{a}=2}^\infty a_{\widehat{a}_1\widehat{a}}A_{\widehat{a}%
}=\alpha _{\widehat{a}_1},\sum\limits_{\widehat{b}=1}^\infty b_{\widehat{b}_1%
\widehat{b}}B_{\widehat{b}}=\beta _{\widehat{b}_1},\sum\limits_{\widehat{c}%
=0}^\infty c_{\widehat{c}_1\widehat{c}}C_{\widehat{c}}=\gamma _{\widehat{c}%
_1},
$$
where $\widehat{a}_1=2,3,\ldots ,\widehat{b}_1=1,2,\ldots ,\widehat{c}%
_1=0,1,\ldots ,$ and there are symmetric brackets
\begin{eqnarray}
\label{symbr}
a_{\widehat{a}_1\widehat{a}} & = & \{\zeta ^{\widehat{a}_1},
\zeta ^{\widehat{a}}\}, \\
b_{\widehat{b}_1\widehat{b}} &= &\{\zeta ^{\widehat{b}_1}p^\alpha
,\zeta ^{\widehat{b}}\triangle _\alpha ^vp_\nu \}, \nonumber \\
c_{\widehat{c}_1\widehat{c}} & = &
\{\zeta ^{\widehat{c}_1}p^\alpha p^\beta ,\zeta ^{\widehat{c}}\triangle
_{\alpha \beta }^{v\mu }p_\nu p_\mu \} \nonumber
\end{eqnarray}
and integrals
\begin{eqnarray}
\alpha _{\widehat{a}_1} & = &
\int \frac{\zeta ^{\widehat{a}_1}
\Xi f_{[0]}}{k_BT\widetilde{n}^2}d\varsigma _p, \label{alpha} \\
 \beta _{\widehat{b}_1} & = &
\int \frac{\zeta ^{\widehat{b}_1}\left( h-p^\alpha U_\alpha \right)
\triangle _{\mu \nu }p^\mu
p^\nu f_{[0]}}{k_BT\widetilde{n}^2}d\varsigma _p, \nonumber \\
\gamma _{\widehat{c}_1} & = &
\int \frac{\zeta ^{\widehat{c}_1}\triangle _{\alpha
\beta }^{v\mu }p_\nu p_\mu p^\alpha p^\beta f_{[0]}}{k_BT\widetilde{n}^2}%
d\varsigma _p. \nonumber
\end{eqnarray}

The lowest approximation is given by the coefficients
$$
A_{\widehat{2}}=\alpha _{\widehat{2}}/a_{\widehat{2}\widehat{2}},B_{\widehat{%
1}}=\beta _{\widehat{1}}/b_{\widehat{1}\widehat{1}},C_{\widehat{0}}=\gamma _{%
\widehat{0}}/c_{\widehat{0}\widehat{0}}.
$$
Introducing these values into (\ref{transpcoef2}) we obtain the first--order
approximations to the transport coefficients
\begin{eqnarray}
\label{transpcoef3}
\lambda &= &-\frac{ck_B\widetilde{n}^2}{n+m-1}
\frac{\beta _{\widehat{1}}^2}{a_{\widehat{2}\widehat{2}}}, \\
\eta & = & \frac{ck_BT\widetilde{n}^2}{\left( n+m\right)
 \left( n+m-1\right) -2}
\frac{\gamma _{\widehat{0}}^2}{c_{\widehat{0}\widehat{0}}}, \nonumber \\
\eta _{(v)} & = & ck_BT\widetilde{n}^2\frac{\alpha _{\widehat{2}}^2}
{a_{\widehat{2}\widehat{2}}}. \nonumber
\end{eqnarray}
The values $\alpha _{\widehat{2}},\beta _{\widehat{1}}$ and $\gamma _{%
\widehat{0}}$ are $\left( n+m-1\right) $--fold integrals which can be
expressed in terms of enthalpy $h$ and temperature $T,$%
\begin{eqnarray} \label{transpcoef4}
\alpha _{\widehat{2}} & = & \frac{n+m+1-\left( n+m-1\right)
\gamma }{ck_BT\widetilde{n}}-\frac{\left( n+m-1\right)
 \gamma }{c\widetilde{n}}
 \underbrace{\longrightarrow}_{T \to \infty}  0,  \\
\beta _{\widehat{1}} & = &
\frac{k_BT}{c\widetilde{n}}\frac{(n+m-1)\gamma }{\gamma-1}
 \underbrace{\longrightarrow}_{T \to \infty}
 \frac{k_BT}{c\widetilde{n}}\left( n+m-1\right) \left(n+m\right) ,
 \nonumber \\
\gamma _{\widehat{0}} & = & \frac{k_BT}{c^2\widetilde{n}}\left( n+m-2\right)
\left( n+m+1\right) h
 \underbrace{\longrightarrow}_{T \to \infty}
 \frac 1{\widetilde{n}}\left( \frac{k_BT}c\right) ^2
\left[ {(n+m)}^2-1\right] \left( n+m\right) . \nonumber
\end{eqnarray}

In order to complete the calculus of the transport coefficients it is
necessary to compute the brackets from the denominators of (\ref{transpcoef3}%
).

\subsection{Brackets from flat to la--spacetimes}

The calculation of brackets is a quite tedious task (see \cite{boisseau})
which should involve the curved spacetime metric and connection. For
simplicity, we consider a background flat spacetime with trivial local
anisotropy. In this case we can apply directly the formulas proved for the
Boltzmann theory in $n_{(a)}$ dimension but with respect to la--frames and
by introducing the locally anisotropic d--connections and d--metrics instead
of their isotropic analogous.

Let us consider a locally adapted to N--connection rest frame, $%
\overrightarrow{e}$ being a unit vector and denote by $\left( \Theta
_{n_{(a)}-1},\Theta _{n_{(a)}-2},...,\Theta _1\right) \,$ the angles of
spherical coordinates with respect to a Cartesian one $\left(
X^1,X^2,...,X^{n_{(a)}}\right) $ chosen as the $X^{n_{(a)}}$--axis to be
along $\overrightarrow{p}_{CM}$ (where $|\overrightarrow{p}_{CM}|=F/%
\widetilde{P},\widetilde{P}=|P^\alpha |,\,$ see (\ref{moller}) (\ref{totmom}%
), and $\overrightarrow{p}_{CM}$ $\overrightarrow{e}$ in the plane of $%
X^{n_{(a)}}$ and $X^{n_{(a)}-1}$ axes. We suppose that the scattering angle
is $\Theta _{n_{(a)}-1}$ and the differential cross section $\left( d\sigma
/d\Omega \right) _{CM}$ is a function only on the module of total energy and
on scattering angle $\Theta =\Theta _{n_{(a)}-1},$%
$$
\left( \frac{d\sigma }{d\Omega }\right) _{CM}=\left( \frac{d\sigma }{d\Omega
}\right) _{CM}\left( \widetilde{P},\Theta \right) .
$$
The results of calculations of symmetric brackets will be expressed into
terms of twofold integrals, which for arbitrary integers
\begin{eqnarray} \label{integral}
& J_{\widehat{r}\widehat{s}}^{\left( \widehat{i},\widehat{j},
\widehat{k}\right) } & \left( \zeta ,\widehat{t},
\widehat{u},\widehat{v},\widehat{w}%
\right)   =     \nu \left( \zeta \right) \left(
\begin{array}{c}
\widehat{r} \\ \widehat{t}
\end{array}
\right) \left(
\begin{array}{c}
\widehat{s} \\ \widehat{u}
\end{array}
\right) \left(
\begin{array}{c}
\widehat{d}/2 \\ \widehat{v}
\end{array}
\right) \frac{2^{\widehat{b}+\widehat{v}-\widehat{d}}}{\sqrt{\pi }}
  \\
& \times & \Gamma\left[ \widehat{b}+\widehat{v}+\frac 12\right]
 F\left( \widehat{t},\widehat{u},\widehat{v}\right)
    \int\limits_{2\zeta }^\infty d\chi
~\chi ^{(n+m-1+\widehat{d}-\widehat{u}-\widehat{b}-
\widehat{v}+\widehat{i})} \nonumber \\
& \times& \left( \frac{\chi ^2}4-
\zeta ^2\right) ^{\left( n+m-2+\widehat{t}+\widehat{u}+\widehat{j}\right)
/2}  K_{\widehat{b}+\widehat{v}-\widehat{h}}\left( \zeta \right)
   \left[ \delta _{0\widehat{w}}-\left(
\begin{array}{c}
\widehat{u} \\ \widehat{w}
\end{array}
\right) \sin ^{\widehat{w}}\Theta \cos ^{\widehat{u}-\widehat{w}}\Theta
\right] \nonumber \\
 &\times& \int\limits_{2\zeta }^\pi d\Theta \sin ^{n+m-1}
\Theta ~\left( \frac{d\sigma }{d\Omega }\right) _{CM}
\left( \widetilde{P},\Theta \right) ,
\nonumber
\end{eqnarray}
where there are introduced the integers
$$
\widehat{d}=\widehat{r}+\widehat{s}-\widehat{t}-\widehat{u}, \widehat{b}%
=\left( \widehat{t}+\widehat{u}+n+m-2\right) /2, \widehat{h}=[1-\left(
-1\right) ^{\widehat{d}}]
$$
and variables $\chi =c/\left( k_BT\right) $ and $\zeta =\widetilde{m}%
c^2/\left( k_BT\right) .$ The factors in (\ref{integral}) are defined
$$
\nu \left( \zeta \right) = \frac{\pi ^{\left( n+m-3\right) /2}}{2\Gamma
\left( (n+m-1)/2\right) \Gamma \left( n+m-3\right) \zeta ^2 \left[ K_{\frac{%
n+m}2}\left( \zeta \right) \right] ^2},
$$
and
\begin{eqnarray}
 & F  \left( \widehat{t},\widehat{u},\widehat{w}\right) &  =
\frac 14\left[ 1+\left(
-1\right) ^{\widehat{w}}\right]
 \left[ 1+\left( -1\right) ^{\widehat{t}+\widehat{u}-\widehat{w}}\right]
  \nonumber \\
 & &\times
 B\left( \frac{n+m-3}2,\frac{\widehat{w}+1}2\right)
  \times
 B\left( \frac{n+m-2+\widehat{w}}2,\frac{\widehat{t}+\widehat{u}-
\widehat{w}+1}2\right) ,
 \nonumber
\end{eqnarray}
where the beta function is given by gamma functions,%
$$
B\left( x,y\right) =\frac{\Gamma \left( x\right) \Gamma \left( y\right) }{%
\Gamma \left( x+y\right) },
$$
and by $\left(
\begin{array}{c}
\widehat{r} \\ \widehat{t}
\end{array}
\right) $ it is denoted the Newton's binomium.

\subsubsection{Scalar type brackets}

The first type of brackets necessary for calculation of transport
coefficients (\ref{transpcoef2}) (see the series approximation (\ref{symbr})
are the so--called the scalar brackets, decomposed into a sum of two
integrals
\begin{equation}
\label{scbr}a_{\widehat{a}_1\widehat{a}}=a_{\widehat{a}_1\widehat{a}%
}^{\prime }+a_{\widehat{a}_1\widehat{a}}^{\prime \prime },
\end{equation}
where $\widehat{a}_1, \widehat{a}=2,3,\ldots ,$%
\begin{eqnarray}
a_{\widehat{a}_1\widehat{a}}^{\prime }&
   = & \widetilde{n}^{-2}\int f_{[0]}\left(
p\right) f_{[0]}\left( p_{[1]}\right)
 \left( \zeta _p\right) ^{\widehat{a}_1}
 \times
 [ \left( \zeta _{p_{[1]}}\right) ^{\widehat{a}}-\left( \zeta
_{p_{[1]}^{\prime }}\right) ^{\widehat{a}}]
 W d\varsigma _pd\varsigma
_{p_{[1]}}d\varsigma _{p^{\prime }}d\varsigma _{p_{[1]}^{\prime }},
 \nonumber  \\
a_{\widehat{a}_1\widehat{a}}^{\prime \prime }&  = &\widetilde{n}^{-2}\int
f_{[0]}\left( p\right) f_{[0]}\left( p_{[1]}\right)
\left( \zeta _p\right) ^{\widehat{a}_1}
  \times [ \left( \zeta _p\right) ^{\widehat{a}}-\left( \zeta
_{p^{\prime }}\right) ^{\widehat{a}}] W d\varsigma _pd\varsigma
_{p_{[1]}}d\varsigma _{p^{\prime }}d\varsigma _{p_{[1]}^{\prime }},
 \nonumber
\end{eqnarray}
with the integration variables
\begin{eqnarray}
\zeta _p & = &
p^\alpha U_\alpha /k_BT,\zeta _{p_{[1]}}=p_{[1]}^\alpha U_\alpha
/k_BT, \nonumber \\
\zeta _{p^{\prime }} &=& p^{\prime \alpha }U_\alpha /k_BT,
\zeta_{p_{[1]}^{\prime }}=p_{[1]}^{\prime \alpha }U_\alpha /k_BT.
 \nonumber
\end{eqnarray}
The explicit calculations of integrals from (\ref{scbr}) (see the method and
basic intermediar formulas in \cite{boisseau}; in this paper we deal with
la--values) give%
\begin{eqnarray}
a_{\widehat{a}_1\widehat{a}}^{\prime } &= &
\sum\limits_{\widehat{t}=0}^{\widehat{a}_1}
\sum\limits_{\widehat{u}=0}^{\widehat{a}}\sum
\limits_{\widehat{v}=0}^{\widehat{d}/2}\sum
\limits_{\widehat{w}=0}^{\widehat{u}}\left( -1\right) ^{\widehat{u}}
J_{\widehat{a}_1\widehat{a}}^{(0,0,0)}\left( \widehat{t},%
\widehat{u},\widehat{v},\widehat{w}\right) , \nonumber \\
a_{\widehat{a}_1\widehat{a}}^{\prime \prime } & = &
\sum\limits_{\widehat{t}=0}^{\widehat{a}_1}
\sum\limits_{\widehat{u}=0}^{\widehat{a}}\sum
\limits_{\widehat{v}=0}^{\widehat{d}/2}\sum
\limits_{\widehat{w}=0}^{\widehat{u}}
J_{\widehat{a}_1\widehat{a}}^{(0,0,0)}\left( \widehat{t},
\widehat{u},\widehat{v},\widehat{w}\right) . \nonumber
\end{eqnarray}
In the lowest approximation, for $\widehat{a}_1=\widehat{a}=2,$ we obtain
$$
a_{\widehat{2}\widehat{2}}=2\left[ J_{\widehat{2}\widehat{2}%
}^{(0,0,0)}\left( \widehat{2},\widehat{2},\widehat{0},\widehat{0}\right) +J_{%
\widehat{2}\widehat{2}}^{(0,0,0)}\left( \widehat{2},\widehat{2},\widehat{0},%
\widehat{2}\right) \right] .
$$
So, we have reduced the scalar brackets to twofold integrals which are
expressed in terms o spherical coordinates with respect to a locally
anisotropic rest frame.

\subsubsection{Vector type brackets}

The vector type brackets from (\ref{transpcoef2}) and (\ref{symbr}) are also
split into two types of integrals
\begin{equation}
\label{vbr}b_{\widehat{b}_1\widehat{b}}=b_{\widehat{b}_1\widehat{b}}^{\prime
}+b_{\widehat{b}_1\widehat{b}}^{\prime \prime }
\end{equation}
where $\widehat{b}_1,\widehat{b}=1,2,\ldots ,$%
\begin{eqnarray}
b_{\widehat{b}_1\widehat{b}}^{\prime } & = &
\widetilde{n}^{-2}\int f_{[0]}\left(
p\right) f_{[0]}\left( p_{[1]}\right) \triangle _{\sigma \varepsilon }\left(
\zeta _p\right) ^{\widehat{b}_1}p^\sigma \nonumber \\
 & & \times
 \left[ p_{[1]}^\varepsilon \left(
\zeta _{p_{[1]}}\right) ^{\widehat{b}}-p_{[1]}^{\prime \varepsilon }\left(
\zeta _{p_{[1]}^{\prime }}\right) ^{\widehat{b}}\right]
  W d\varsigma_pd
\varsigma _{p_{[1]}}d\varsigma _{p^{\prime }}d\varsigma _{p_{[1]}^{\prime
}}, \nonumber \\
b_{\widehat{b}_1\widehat{b}}^{\prime \prime } & = & \widetilde{n}^{-2}\int
f_{[0]}\left( p\right) f_{[0]}\left( p_{[1]}\right) \triangle _{\sigma
\varepsilon }\left( \zeta _p\right) ^{\widehat{b}_1}p^\sigma  \nonumber \\
 & & \times \left[
p^\varepsilon \left( \zeta _p\right) ^{\widehat{b}}-p^{\prime \varepsilon
}\left( \zeta _{p^{\prime }}\right) ^{\widehat{b}}\right]
  W d\varsigma_pd\varsigma _{p_{[1]}}
d\varsigma _{p^{\prime }}d\varsigma _{p_{[1]}^{\prime}}. \nonumber
\end{eqnarray}
A tedious calculus similar to that presented in \cite{boisseau} implies
further decompositions of coefficients and their representation as
$$
b_{\widehat{b}_1\widehat{b}}^{\prime }=\sum_{\left( i\right) =1}^3b_{(i)%
\widehat{b}_1\widehat{b}}^{\prime } \mbox{ and } b_{\widehat{b}_1\widehat{b}%
}^{\prime \prime }=\sum_{\left( i\right) =1}^3b_{(i)\widehat{b}_1\widehat{b}%
}^{\prime \prime },
$$
with corresponding sums%
\begin{eqnarray}
b_{(1)\widehat{b}_1\widehat{b}}^{\prime } & = &
\frac 14\left( \frac{k_BT}c\right)^2  \times
\sum\limits_{\widehat{t}=0}^{\widehat{b}_1}
\sum\limits_{\widehat{u}=0}^{\widehat{b}}
\sum\limits_{\widehat{v}=0}^{\widehat{d}/2}
\sum\limits_{\widehat{w}=0}^{\widehat{u}}
\left( -1\right) ^{\widehat{u}}J_{\widehat{a}_1\widehat{a}}^{(2,0,0)}
\left( \widehat{t},\widehat{u},\widehat{v},\widehat{w}\right) ,
\nonumber \\
b_{(2)\widehat{b}_1\widehat{b}}^{\prime } & = &\left( \frac{k_BT}c\right)^2
  \sum\limits_{\widehat{t}=0}^{\widehat{b}_1}
\sum\limits_{\widehat{u}=0}^{\widehat{b}}
\sum\limits_{\widehat{v}=0}^{\widehat{d}/2}
\sum\limits_{\widehat{w}=0}^{\widehat{u}}
\left( -1\right) ^{\widehat{u}}
J_{\widehat{a}_1\widehat{a}}^{(0,2,1)}
\left( \widehat{t},\widehat{u},\widehat{v},\widehat{w}\right) ,
\nonumber \\
b_{(2)\widehat{b}_1\widehat{b}}^{\prime } & = & -\left( \frac{k_BT}c\right)^2
 \sum\limits_{\widehat{t}=0}^{\widehat{b}_1}
\sum\limits_{\widehat{u}=0}^{\widehat{b}}
\sum\limits_{\widehat{v}=0}^{\widehat{d}/2}
\sum\limits_{\widehat{w}=0}^{\widehat{u}}
\left( -1\right) ^{\widehat{u}}J
_{\widehat{a}_1+1\widehat{a}+1}^{(0,2,1)}
\left( \widehat{t},\widehat{u},\widehat{v},\widehat{w}\right);
\nonumber
\end{eqnarray}
with corresponding sum expressions for $b_{(i)\widehat{b}_1\widehat{b}%
}^{\prime \prime }$ by omitting the factor $\left( -1\right) ^{\widehat{u}}.$

In the lowest approximation (for $\widehat{b}_1=\widehat{b}=1)$ we have
\begin{eqnarray}
b_{\widehat{1}\widehat{1}} & = &-2\left( \frac{k_BT}c\right) ^2
 \times [ J_{\widehat{1}\widehat{1}}^{(0,2,1)}
\left( \widehat{1},\widehat{1},\widehat{0},\widehat{0}\right) \nonumber \\
 & & + J_{\widehat{2}\widehat{2}}^{(0,0,0)}\left( \widehat{2},\widehat{2},%
\widehat{0},\widehat{0}\right) +J_{\widehat{2}\widehat{2}}^{(0,0,0)}\left(
\widehat{2},\widehat{2},\widehat{0},\widehat{2}\right) ] .
 \nonumber
 \end{eqnarray}
Here should be noted that in general $T=T\left( u\right) $ is a function on
la--spacetime coordinates.

\subsubsection{Tensor type brackets}

In a similar fashion as for scalar and vector type symmetric brackets from (%
\ref{transpcoef2}) and (\ref{symbr}) one holds the decomposition
\begin{equation}
\label{tbr}c_{\widehat{c}_1\widehat{c}}=c_{\widehat{c}_1\widehat{c}}^{\prime
}+c_{\widehat{c}_1\widehat{c}}^{\prime \prime }
\end{equation}
where $\widehat{c}_1,\widehat{c}=1,2,\ldots ,$%
\begin{eqnarray}
 & c_{\widehat{c}_1\widehat{c}}^{\prime } & =
\widetilde{n}^{-2}\int f_{[0]}\left(
p\right) f_{[0]}\left( p_{[1]}\right) \triangle _{\sigma \varepsilon }^{\nu
\mu }\left( \zeta _p\right) ^{\widehat{c}_1}p_\nu p_\mu  \nonumber \\
  & & \times
[ p_{[1]}^\sigma p_{[1]}^\varepsilon
\left( \zeta _{p_{[1]}}\right) ^{\widehat{c}}-
p_{[1]}^{\prime \sigma }p_{[1]}^{\prime \varepsilon }\left( \zeta
_{p_{[1]}^{\prime }}\right) ^{\widehat{c}}]
 W d\varsigma _pd\varsigma _{p_{[1]}}d\varsigma _{p^{\prime
}}d\varsigma _{p_{[1]}^{\prime }},  \nonumber \\
&c_{\widehat{c}_1\widehat{c}}^{\prime \prime } &= \widetilde{n}^{-2}\int
f_{[0]}\left( p\right) f_{[0]}\left( p_{[1]}\right) \triangle _{\sigma
\varepsilon }^{\nu \mu }\left( \zeta _p\right) ^{\widehat{c}_1}p_\nu p_\mu
 \nonumber \\
 & & \times
[ p^\sigma p^\varepsilon \left( \zeta _{p_{[1]}}\right) ^{\widehat{c}}-
p^{\prime \sigma }p^{\prime \varepsilon }\left( \zeta _{p_{[1]}^{\prime
}}\right) ^{\widehat{c}}]  W
d\varsigma _pd\varsigma _{p_{[1]}}d\varsigma _{p^{\prime
}}d\varsigma _{p_{[1]}^{\prime }}.  \nonumber
\end{eqnarray}
For tensor like brackets we have to consider sums on nine terms
$$
c_{\widehat{c}_1\widehat{c}}^{\prime }=\sum_{(s)=1}^9c_{(s)\widehat{c}_1%
\widehat{c}}^{\prime } \mbox{ and } c_{\widehat{c}_1\widehat{c}}^{\prime
\prime }= \sum_{(s)=1}^9c_{(s)\widehat{c}_1\widehat{c}}^{\prime \prime }.
$$
These terms are \cite{boisseau} (with that exception that we have
dependencies on the number of anisotropic variables and $T=T\left( u\right) $
is a function on la--spacetime coordinates)%
\begin{eqnarray}
c_{(1)\widehat{c}_1\widehat{c}}^{\prime } & = &
 \frac 1{16}\left( \frac{k_BT}c\right) ^4
\sum\limits_{\widehat{t}=0}^{\widehat{c}_1}
\sum\limits_{\widehat{u}=0}^{\widehat{c}}
\sum\limits_{\widehat{v}=0}^{\widehat{d}/2}
\sum\limits_{\widehat{w}=0}^{\widehat{u}}
\left( -1\right) ^{\widehat{u}}J_{\widehat{c}_1\widehat{c}}^{(4,0,0)}
\left( \widehat{t},\widehat{u},\widehat{v},\widehat{w}%
\right) , \nonumber\\
c_{(2)\widehat{c}_1\widehat{c}}^{\prime } & = &\left( \frac{k_BT}c\right)^4
\sum\limits_{\widehat{t}=0}^{\widehat{c}_1}
\sum\limits_{\widehat{u}=0}^{\widehat{c}}
\sum\limits_{\widehat{v}=0}^{\widehat{d}/2}
\sum\limits_{\widehat{w}=0}^{\widehat{u}}
\left( -1\right) ^{\widehat{u}}J_{\widehat{c}_1\widehat{c}}^{(0,4,2)}
\left( \widehat{t},\widehat{u},\widehat{v},\widehat{w}\right) ,
\nonumber \\
c_{(3)\widehat{c}_1\widehat{c}}^{\prime } & = &
\frac{n+m-2}{n+m-1}\left( \frac{k_BT}c\right) ^4
 \sum\limits_{\widehat{t}=0}^{\widehat{c}_1}
\sum\limits_{\widehat{u}=0}^{\widehat{c}}
\sum\limits_{\widehat{v}=0}^{\widehat{d}/2}
\sum\limits_{\widehat{w}=0}^{\widehat{u}}
\left( -1\right) ^{\widehat{u}}
J_{\widehat{c}_1+2\widehat{c}+2}^{(0,0,0)}\left( \widehat{t},\widehat{u},%
\widehat{v},\widehat{w}\right) , \nonumber \\
c_{(4)\widehat{c}_1\widehat{c}}^{\prime } & = &
\frac 12\left( \frac{k_BT}c\right)^4
 \sum\limits_{\widehat{t}=0}^{\widehat{c}_1}
\sum\limits_{\widehat{u}=0}^{\widehat{c}}
\sum\limits_{\widehat{v}=0}^{\widehat{d}/2}
\sum\limits_{\widehat{w}=0}^{\widehat{u}}
\left( -1\right) ^{\widehat{u}}
J_{\widehat{c}_1\widehat{c}}^{(2,2,1)}
\left( \widehat{t},\widehat{u},\widehat{v},\widehat{w}\right) ,
\nonumber \\
c_{(5)\widehat{c}_1\widehat{c}}^{\prime } & = &-2\left( \frac{k_BT}c\right)^4
  \sum\limits_{\widehat{t}=0}^{\widehat{c}_1}
\sum\limits_{\widehat{u}=0}^{\widehat{c}}
\sum\limits_{\widehat{v}=0}^{\widehat{d}/2}
\sum\limits_{\widehat{w}=0}^{\widehat{u}}
\left( -1\right) ^{\widehat{u}}J_{\widehat{c}_1+1\widehat{c}+1}^{(0,2,1)}
\left( \widehat{t},\widehat{u},\widehat{v},\widehat{w}\right),
 \nonumber \\
c_{(6)\widehat{c}_1\widehat{c}}^{\prime } & = &-\frac 12
\left( \frac{k_BT}c\right) ^4
 \sum\limits_{\widehat{t}=0}^{\widehat{c}_1}
\sum\limits_{\widehat{u}=0}^{\widehat{c}}
\sum\limits_{\widehat{v}=0}^{\widehat{d}/2}
\sum\limits_{\widehat{w}=0}^{\widehat{u}}
\left( -1\right) ^{\widehat{u}}
J_{\widehat{c}_1+1\widehat{c}+1}^{(2,0,0)}
\left( \widehat{t},\widehat{u},\widehat{v},\widehat{w}\right) , \nonumber \\
c_{(7)\widehat{c}_1\widehat{c}}^{\prime } & = &-\frac{\zeta ^4}{n+m-1}\left(
\frac{k_BT}c\right) ^4
 \sum\limits_{\widehat{t}=0}^{\widehat{c}_1}
\sum\limits_{\widehat{u}=0}^{\widehat{c}}
\sum\limits_{\widehat{v}=0}^{\widehat{d}/2}
\sum\limits_{\widehat{w}=0}^{\widehat{u}}
\left( -1\right) ^{\widehat{u}}
J_{\widehat{c}_1\widehat{c}}^{(0,0,0)}\left( \widehat{t},%
\widehat{u},\widehat{v},\widehat{w}\right) , \nonumber \\
c_{(8)\widehat{c}_1\widehat{c}}^{\prime } & = &
\frac{\zeta ^4}{n+m-1}\left( \frac{k_BT}c\right) ^4
 \sum\limits_{\widehat{t}=0}^{\widehat{c}_1}
\sum\limits_{\widehat{u}=0}^{\widehat{c}}
\sum\limits_{\widehat{v}=0}^{\widehat{d}/2}
\sum\limits_{\widehat{w}=0}^{\widehat{u}}
\left( -1\right) ^{\widehat{u}}
J_{\widehat{c}_1+2\widehat{c}}^{(0,0,0)}\left( \widehat{t},\widehat{u},%
\widehat{v},\widehat{w}\right) ,  \nonumber \\
c_{(9)\widehat{c}_1\widehat{c}}^{\prime } & = &
\frac{\zeta ^4}{n+m-1}\left( \frac{k_BT}c\right) ^4
 \sum\limits_{\widehat{t}=0}^{\widehat{c}_1}
\sum\limits_{\widehat{u}=0}^{\widehat{c}}
\sum\limits_{\widehat{v}=0}^{\widehat{d}/2}
\sum\limits_{\widehat{w}=0}^{\widehat{u}}
\left( -1\right) ^{\widehat{u}}
J_{\widehat{c}_1\widehat{c}+2}^{(0,0,0)}
\left( \widehat{t},\widehat{u},%
\widehat{v},\widehat{w}\right) . \nonumber
\end{eqnarray}
The twice primed values $c_{(s)\widehat{c}_1\widehat{c}}^{\prime \prime }$
are given by similar sums by omitting the factor $\left( -1\right) ^{%
\widehat{u}}$ and by changing into $c_{(4)\widehat{c}_1\widehat{c}}^{\prime
\prime }$ and $c_{(5)\widehat{c}_1\widehat{c}}^{\prime \prime }$ the overall
signs.

In the lower approximation, for $\widehat{c}_1=\widehat{c}=0,$ one holds%
\begin{eqnarray}
&c_{\widehat{0}\widehat{0}} & =
2\left( \frac{k_BT}c\right) ^4\{J_{\widehat{0}\widehat{0}}^{(0,4,2)}
( \widehat{0},\widehat{0},\widehat{0},\widehat{0}) +
2J_{\widehat{1}\widehat{1}}^{(0,2,1)} ( \widehat{1},\widehat{1},%
\widehat{0},\widehat{0})  \nonumber \\
& & +  \frac{n+m-2}{n+m-1} [ J_{\widehat{2}\widehat{2}}^{(0,0,0)}
(\widehat{2},\widehat{2},\widehat{0},\widehat{0}) +
 J_{\widehat{2}\widehat{2}}^{(0,0,0)}
( \widehat{2},\widehat{2},\widehat{0},\widehat{2}) ] \}. \nonumber
\end{eqnarray}

\subsection{Locally anisotropic transport coefficients for a class of cross
sections}

For astrophysical applications we can substitute
\begin{equation}
\label{lepton}\left( \frac{d\sigma }{d\Omega }\right) _{CM}=\xi P^r,
\end{equation}
with some scalar factor $\xi $ (with or without dimension) and $r$ being a
positive or negative number into (\ref{integral}). In this subsection we put
$\zeta =\widetilde{m}c^2/\left( k_BT\right) $ for high values of $T.$ The
chosen type of differential cross section (\ref{lepton}) is used, for
instance, for calculations of neutrino--neutrino scattering (when $r=2$ and $%
\xi $ is connected with the weak coupling constant). In the first
approximation the symmetric brackets (\ref{symbr}) are
\begin{eqnarray} \label{symbrfa}
&a_{\widehat{2}\widehat{2}} &=
\xi \left( \frac{2k_BT}c\right) ^r
\frac{\pi ^{(n+m-1)/2} \Gamma \left[n+m+1+r/2\right] \Gamma
\left[ \left( n+m+r\right)/2\right] }{\Gamma
\left[ n+m-2\right] \Gamma \left[ \left( n+m\right)
/2\right] \Gamma \left[
\left( n+m+1\right) /2\right] },  \\
 & b_{\widehat{1}\widehat{1}} & =
 - \left( \frac{k_BT}c\right) ^2\frac{n+m+2+r}2a_{\widehat{2}\widehat{2}},
\nonumber \\
& c_{\widehat{0}\widehat{0}} & =
\left( \frac{k_BT} c\right) ^2
\left[ \frac{\left(
n+m+r\right) \left( n+m+4+r\right) }2+\frac{n+m-2}{n+m-1}\right]
 a_{\widehat{2}\widehat{2}}. \nonumber
\end{eqnarray}

Putting these values into (\ref{transpcoef4}) we obtain the locally
anisotropic variant of transport coefficients in the first approximation,
when $\eta _{(v)}\simeq 0$ but with nonzero
\begin{eqnarray}
\lambda  & = &
\frac{2k_Bc}\xi \left( \frac c{2k_BT}\right) ^r
\frac{\left( n+m\right) ^2\left( n+m-1\right) }{n+m+2+r}
 \label{frl}  \\
&{}& \times \frac{\Gamma \left[ n+m-2\right] \Gamma \left[
\left(n+m\right)/2\right]
 \Gamma \left[ \left( n+m+1\right) /2\right] }{\Gamma
\left[n+m+1+r/2\right] \Gamma \left[ \left( n+m+r\right) /2\right] },
\nonumber \end{eqnarray}
and
\begin{eqnarray}
\eta  & = & \frac 1\xi \left( \frac c{2k_BT}\right) ^{r-1}\frac 1{\pi
^{(n+m-1)/2}\left[ \left( n+m\right) ^2-n-m-2\right] } \nonumber \\
 &\times & \frac{ \left[ (n+m-1)^2-1\right] \left(n+m+1\right) ^2}
 {\left[  \left( n+m+r\right) \left(n+m+4+r\right) +
 2 { \left( n+m-2\right) \over \left( n+m-1\right) }
\right] } \label{fre} \\
 &\times &
\frac{\Gamma \left[ n+m-2\right] \Gamma \left[ \left( n+m\right)
/2\right] \Gamma \left[ \left( n+m+1\right) /2\right] }{\Gamma \left[
n+m+1+r/2\right] \Gamma \left[ \left( n+m+r\right) /2\right] }.
 \nonumber
\end{eqnarray}

These formulas present a locally anisotropic generalization of the Boisseau
and van Leeuwen \cite{boisseau} results.

\section{Transport Theory in Curved Spaces with Rotation Ellipsoidal Horizons
}

After having established in a general way the scheme for calculation of
transport coefficients on la--spacetimes, we now specify an example for a
four dimensional static metric (being a solution of Einstein equations in
both general relativity and la--gravity) with the event horizon described by
a hypersurface of rotation ellipsoid \cite{v6} (see the Appendix). We note
that in this case $n=3,m=1$ and $n_{(a)}=4.$

The main formulas for kinetic and thermodynamic variables from the previous
Sections 3--6 were based on spherical symmetry of $n_{(a)}-1$ dimensional
volume (\ref{sphvol}) and spherical integral (\ref{sphint}). For the
rotation ellipsoid we have to modify the volume's formula by introducing the
ellipsoidal dependence
$$
V_{n_{(a)}-1} = \frac{\left( n_{(a)}-1\right) \pi ^{(n_{(a)}-1)/2}}{\Gamma
\left[ (n_{(a)}-1)/2\right] }\rho \left( \widetilde{\theta }\right)
\underbrace { = }_{n_{(a)}=4} \frac{3\pi ^{3/2}}{\Gamma \left[ 3/2\right] }%
\rho \left( \widetilde{\theta }\right)
$$
were
$$
\rho \left( \widetilde{\theta }\right) =\frac{\rho _{(0)}}{1-\varepsilon
\cos \widetilde{\theta }}
$$
is the parametric formula of an ellipse with constant parameter $\rho
_{(0)}, $ angle variable $\widetilde{\theta}$ and eccentricity $\varepsilon
=1/\sigma <1$ is defined by the axes of rotation ellipsoid (see the formula (%
\ref{ele4}) from Appendix).

If in the case of spherical symmetry $4\int\limits_0^{\pi /2}d \widetilde{%
\theta} =2\pi , $ the ellipse deformation gives the result
$$
4\int\limits_0^{\pi /2}\frac{d \widetilde{\theta} } {1-\varepsilon \cos
\widetilde{ \theta} }=\frac 8{\sqrt{1-\varepsilon ^2}}\arctan \sqrt{\frac{%
1-\varepsilon }{1+\varepsilon }}.
$$
So, performing integrations on solid angles in spaces with rotational
ellipsoid symmetry we can use the same formulas as for spherical symmetry
but multiplied on
$$
q_{(\varepsilon )}=\frac 4{\pi \sqrt{1-\varepsilon ^2}}\arctan \sqrt{\frac{%
1-\varepsilon }{1+\varepsilon }}.
$$
For instance, the integral (\ref{sphint}) transforms
$$
\int d\Omega \rightarrow \int d\Omega _{(\varepsilon )} = q_{(\varepsilon )}%
\frac{\left( n_{(a)}-1\right) \pi ^{(n_{(a)}-1)/2}}{\Gamma \left[
(n_{(a)}-1)/2\right] } \underbrace{ = }_{n_{(a)}=4} q_{(\varepsilon )}\frac{%
3\pi ^{3/2}}{\Gamma \left[ 3/2\right] }.
$$

The formulas for the particle density (\ref{parnumvol}) and energy density (%
\ref{edensbess}) of point particles must be multiplied on $q_{(\varepsilon
)},$
$$
\widetilde{n}\rightarrow \widetilde{n}_{(\varepsilon )}=q_{(\varepsilon )}
\widetilde{n} \mbox{ and  } \widetilde{\varepsilon }\rightarrow \widetilde{%
\varepsilon }_{(\varepsilon )}=q_{(\varepsilon )}\widetilde{\varepsilon },
$$
but the energy per particle will remain constant. We also have to modify the
formula for pressure (\ref{pressure}), been proportional to the particle
density, but consider unchanged the averaged enthalpy (\ref{enthalpyp}). The
entropy per particle (\ref{entropy}) and chemical potential (\ref{mupot})
depends explicitly on $q_{(\varepsilon )}$--factor because theirs formulas
were derived by using the particle density $\widetilde{n}_{(\varepsilon )}.$
Here we note that all proved formulas depends on $m(=1$, in this Section)
anisotropic parameters and on volume element determined by d--metric. In the
first approximation of transport coefficients we could chose a locally
isotropic background but introducing the factor $q_{(\varepsilon )}$ and
taking into account the dependence on anisotropic dimension.

Putting $\varepsilon$--corrections into (\ref{transpcoef3}) we obtain the
first--order approximations to the transport coefficients in a la--spacetime
with the symmetry of rotation ellipsoid
\begin{eqnarray}
\label{transpcoef3el}
\lambda &= &-  \frac{ck_B\widetilde{n}^2_{(\varepsilon)}}{3}
\frac{\beta _{\widehat{1}}^2}{a_{\widehat{2}\widehat{2}}}, \\
\eta & = &  \frac{ck_BT\widetilde{n}^2_{(\varepsilon)}}{10}
\frac{\gamma _{\widehat{0}}^2}{c_{\widehat{0}\widehat{0}}}, \nonumber \\
\eta _{(v)} & = & ck_BT\widetilde{n}^2_{(\varepsilon)}
\frac{\alpha _{\widehat{2}}^2}
{a_{\widehat{2}\widehat{2}}}. \nonumber
\end{eqnarray}
The values $\alpha _{\widehat{2}},\beta _{\widehat{1}}$ and $\gamma _{%
\widehat{0}}$ are $3$--fold integrals expressed in terms of enthalpy $h$ and
temperature $T$ and have the limits
\begin{eqnarray}
\alpha _{\widehat{2}} & = & \frac{5
\gamma }{ck_BT\widetilde{n}_{(\varepsilon)}}-
  \frac{ 3 \gamma }{c\widetilde{n}_{(\varepsilon)}}
\underbrace{\longrightarrow}_{T \to \infty}  0, \label{alphael} \\
\beta _{\widehat{1}} & = &
\frac{k_BT}{c\widetilde{n}_{(\varepsilon)}}\frac{3 \gamma }{\gamma-1}
 \underbrace{\longrightarrow}_{T \to \infty}
 12 \frac{k_BT}{c\widetilde{n}_{(\varepsilon)}}, \nonumber \\
\gamma _{\widehat{0}} & = & \frac{k_BT}{c^2
\widetilde{n}_{(\varepsilon)}} 10 h
   \underbrace{\longrightarrow}_{T \to \infty}
 60 \frac 1{\widetilde{n}_{(\varepsilon)}}
\left( \frac{k_BT}c\right) ^2. \nonumber
\end{eqnarray}

In the next step we compute the $q_{(\varepsilon)}$--deformations of
brackets from the denominators of (\ref{transpcoef3el}). Because the squares
of $\alpha , \beta$ and $\gamma$ coefficients from (\ref{alpha}) are
proportional to deformations of $\widetilde{n}^{-2}$ (this conclusion
follows from the formulas (\ref{symbrfa}) in the $T \to \infty$ limit, see (%
\ref{alphael})) and the scalar (\ref{scbr}), vector (\ref{vbr}) and tensor (%
\ref{tbr}) type brackets do not change under $q_{(\varepsilon)}$%
--deformations (see (\ref{symbrfa})) we conclude that the transport
coefficients (\ref{transpcoef3}) do not contain the factor $%
q_{(\varepsilon)} $ but depends only on the number $m$ of anisotropic
dimensions. This conclusion is true only in the first approximation and for
locally isotropic backgrounds. In consequence, the final formulas for the
transport coefficient, see (\ref{frl}) and (\ref{fre}), in a (3+1) locally
anisotropic spacetime with rotation ellipsoid symmetry are
\begin{eqnarray}
\eta _{(v)} &\simeq& 0 \nonumber \\
\lambda  & = & \frac{2k_Bc}\xi \left( \frac c{2k_BT}\right) ^r
\frac{48}{6+r}
 \frac{{\Gamma \left[ 2\right]}^2 \Gamma \left[ 5/2\right] }
{\Gamma\left[5+r/2\right] \Gamma \left[ \left( 4+r\right) /2\right] },
\nonumber \\
\eta  & = & \frac 1\xi \left( \frac c{2k_BT}\right) ^{r-1}
\frac{20}{\pi^{3/2} [(4+r)(8+r) +4/3]}
  \frac{{\Gamma [2]}^2 \Gamma \left[5/2\right] }
 {\Gamma \left[5+r/2\right] \Gamma \left[ \left( 4+r\right) /2\right] }.
 \nonumber
\end{eqnarray}
This is consistent with the fact that we chose as an example a static metric
with a local anisotropy that does not cause drastic changings in the
structure of transport coefficients. Nevertheless, there are $%
q_{(\varepsilon)}$--deformations of such values as the density of particles,
kinetic potential and entropy which reflects modifications of kinetic and
thermodynamic processes even by static spacetime local anisotropies.

\section{Concluding Remarks}

The formulation of Einstein's theory of relativity with respect to
anholonomic frames raises a number of questions concerning locally
anisotropic field interactions and kinetic and thermodynamic effects.

We argue that spacetime local anisotropy (la) can be modeled by applying the
Cartan's moving frame method \cite{cartan1} with associated nonlinear
connection (N--connection) structures. A remarkable fact is that this
approach allows a unified treatment of various type of theories with generic
local anisotropy like generalized Finsler like gravities, of standard
Kaluza--Klein models with nontrivial compactifications (modelled by
N--connection structures), of standard general relativity with anholonomic
frames and even of low dimensional models with distinguished anisotropic
parameters. We have shown a relationship between a subclass of Finsler like
metrics with (pseudo) Riemannian ones being solutions of Einstein equations.

This paper has provided a generalization of relativistic kinetics and
nonequilibrium thermodynamics in order to be included possible spacetime
local anisotropies. It should not be considered as a work on definitions of
some sophisticated theories on Finsler like spaces but developing an approach
to kinetics and thermodynamics in (pseudo) Riemannian spacetimes of
arbitrary dimension provided with anholonomic frame structures. The item of
introducing into consideration of generalized Finsler and Kaluza--Klein
spaces was imposed by the facts that one has recently constructed (see
Appendix and \cite{v6} solutions of the Einstein equations with generic
anisotropy, of Finsler and another type, like ellipsoidal static black
holes, black tora, anisotropic solitonic backgrounds and so on) and that in
the low energy limits of string theories various classes of generalized
Finsler--Kaluza--Klein metrics could be obtained alternatively to the well
known (pseudo) Riemannian ones \cite{v2}. By applying the moving frame
method it is possible to elaborate a general schema for defining of physical
values, basic equations and approximated calculations of kinetic and
thermodynamic values with respect to la--frames in all type of the mentioned
theories.

The crucial ingredient in definition of collisionless relativistic locally
anisotropic kinetic equation was the extension of the moving frame method to
the space of supporting elements $(u^\alpha , p^\beta )$ provided with an
induced higher order anisotropic structure. The former Cartan--Vlasov
approach \cite{cartan,vlas}, proposing a variant of statistical kinetic
theory on curved phase spaces provided with Finsler like and Cartan
N--connection structures, was self--consistently modified for both type of
locally isotropic (the Einstein theory) and an\-isot\-rop\-ic (generalized
Finsler--Kaluza--Klein) spacetimes with N--con\-nec\-ti\-on struc\-tures
in\-duced by lo\-cal an\-holo\-nomic fra\-me or via reductions from higher
dimensions in (super) string or (super) gravity theories. The physics of
pair collisions in la--spacetimes was examined by introducing on the space
of supporting elements of (correspondingly adapted to the N--connection
structure) integral of collisions, differential cross--sections and velocity
of transitions.

Despite all the complexities of definition of equilibrium states with
generic an\-isot\-ro\-py it is possible a rigorous definition of local
equilibrium particle distribution functions by fixing some anholonomic
frames of reference adapted to the N--connection structure. The basic
kinetic and thermodynamic values such as particle density, average energy
and pressure, enthalpy, specific heats and entropy are derived via
integrations on volume elements determined by metric components given with
respect to la--bases. In the low and high energy limits the formulas reflect
explicit dependencies on the number of anisotropic dimensions as well on
anisotropic deformations of spacetime metric and linear connection.

One can linearize the transport equations and prove the linear laws for
locally anisotropic non--equilibrium thermodynamics. It has also established
a general scheme for calculation of transport coefficients (the heat
conductivity, the shear viscosity and the volume viscosity) in
la--spacetimes. An explicit computation of such values was performed for a
metric with rotation ellipsoidal event horizon (an example of spacetime with
static local anisotropy), recently found as a new solution of the Einstein
equations.

Our overall conclusion is that in order to obtain a self--consistent
formulation of the locally anisotropic kinetic and thermodynamic theory in
curved spacetimes and calculation of basic physical values we must consider
moving frames with correspondingly adapted nonlinear connection structures.

\vskip0.2cm {\bf Aknowledgments}

The author want to thank Professors G. Neugebauer and H. Dehnen for
hospitality and support of his participation at Journees Relativistes 99 at
Weimar and visit at Konstantz University, Germany, where the bulk of results
were communicated.

\section*{Appendix:\ A Locally Anisotropic Solution of Einstein's Equations}

Before presenting an explicit construction \cite{v6} of a four dimensional
solution with local anisotropy of the Einstein equations (\ref{einsteq1}) we
briefly review the properties of four dimensional metrics which transforms
into (2+2) or (3+1) anisotropic d--metrics (we note that this is not a
(space + time) but a (isotropic+anisotropic) decomposition of coordinates)
by transitions to correspondingly defined anholonomic bases of tetrads
(vierbeins).

Let us consider a four dimensional (in brief, 4D) spacetime $V^{(2+2)}$
(with two isotropic plus two anisotropic local coordinates) provided with a
metric of signature (+,+,--,+) parametrized by a symmetric matrix of type
\begin{equation}
\label{ansatz2}\left[
\begin{array}{cccc}
g_1+q_1{}^2h_3+n_1{}^2h_4 & 0 & q_1h_3 & n_1h_4 \\
0 & g_2+q_2{}^2h_3+n_2{}^2h_4 & q_2h_3 & n_2h_4 \\
q_1h_3 & q_2h_3 & h_3 & 0 \\
n_1h_4 & n_2h_4 & 0 & h_4
\end{array}
\right]
\end{equation}
with coefficients being some functions
$$
g_i=g_i(x^j),q_i=q_i(x^j,t),n_i=n_i(x^j,t),h_a=h_a(x^j,t)
$$
of necessary smo\-oth class. With respect to a la--basis (\ref{ddif}) the
ansatz (\ref{ansatz2}) results in diagonal $2\times 2$ h-- and v--metrics
for a d--metric (\ref{dmetric}) (for simplicity, we shall consider only
diagonal 2D nondegenerated metrics because for such dimensions every
symmetric matrix can be diagonalized).

An equivalent diagonal d--metric of type (\ref{dmetric}) is obtained for the
associated N--connection with the coefficients being functions on three
coordinates $(x^i,z),$
\begin{eqnarray} N_1^3&=&q_1(x^i,z),\ N_2^3=q_2(x^i,z), \label{ncoef} \\
N_1^4&=&n_1(x^i,z),\ N_2^4=n_2(x^i,z). \nonumber \end{eqnarray}
For simplicity, we shall use brief denotations of partial derivatives, like $%
\dot a$ $=\partial a/\partial x^1,$ $a^{\prime }=\partial a/\partial x^2,$ $%
a^{*}=\partial a/\partial z$ $\dot a^{\prime }$ $=\partial ^2a/\partial
x^1\partial x^2,$ $a^{**}=\partial ^2a/\partial z\partial z.$

The non--trivial components of the Ricci d--tensor (\ref{dricci}) (for the
ansatz \ref{ansatz2}) when $R_1^1=R_2^2$ and $S_3^3=S_4^4,$ are computed
\begin{eqnarray}
&R_1^1&=R_2^2=\frac 1{2g_1g_2} [-(g_1^{^{\prime \prime }}+{\ddot g}_2)+
\frac 1{2g_2}\left( {\dot g}_2^2+g_1^{\prime }g_2^{\prime }\right) +
\frac 1{2g_1}\left( g_1^{\prime \ 2}+%
\dot g_1\dot g_2\right) ]; \label{ricci1} \\
\label{ricci2} &S_3^3&=S_4^4=
\frac 1{h_3h_4}[-h_4^{**}+\frac 1{2h_4}(h_4^{*})^2+%
\frac 1{2h_3}h_3^{*}h_4^{*}]; \\
&P_{31}&=\frac{q_1}2[\left( \frac{h_3^{*}}{h_3}\right) ^2-
\frac{h_3^{**}}{h_3}+%
\frac{h_4^{*}}{2h_4^{\ 2}}-\frac{h_3^{*}h_4^{*}}{2h_3h_4}]
+\frac 1{2h_4}[\frac{\dot h_4}{2h_4}h_4^{*}-\dot h_4^{*}+ %
\frac{\dot h_3}{2h_3}h_4^{*}],  \label{ricci3} \\
&{}& \nonumber \\
&P_{32}&=\frac{q_2}2[\left( \frac{h_3^{*}}{h_3}\right) ^2-
\frac{h_3^{**}}{h_3}+%
\frac{h_4^{*}}{2h_4^{\ 2}}-\frac{h_3^{*}h_4^{*}}{2h_3h_4}]
+\frac 1{2h_4}[\frac{h_4^{\prime }}{2h_4}h_4^{*}-h_4^{\prime \ *}+ %
\frac{h_3^{\prime }}{2h_3}h_4^{*}];  \nonumber \\
&{}&   \nonumber \\
& P_{41}&=-\frac{h_4}{2h_3}n_1^{**} +
\frac{1}{4h_3}(\frac{h_4}{h_3} h^*_3 - 3 h^*_4) n^*_1 , \label{ricci4} \\
& P_{42}&= -\frac{h_4}{2h_3}n_2^{**} +
\frac{1}{4h_3}(\frac{h_4}{h_3} h^*_3 - 3 h^*_4) n^*_2. \nonumber
\end{eqnarray}

The curvature scalar $\overleftarrow{R}$ (\ref{dscalar}) is defined by the
sum of two non-trivial components $\widehat{R}=2R_1^1$ and $S=2S_3^3.$

The system of Einstein equations (\ref{einsteq2}) transforms into
\begin{eqnarray}
R_1^1&=&-\kappa \Upsilon _3^3=-\kappa \Upsilon _4^4,
\label{einsteq3a} \\
S_3^3&=&-\kappa \Upsilon _1^1=-\kappa \Upsilon _2^2, \label{einsteq3b}\\
P_{3i}&=& \kappa \Upsilon _{3i}, \label{einsteq3c} \\
P_{4i}&=& \kappa \Upsilon _{4i}, \label{einsteq3d}
\end{eqnarray}
where the values of $R_1^1,S_3^3,P_{ai},$ are taken respectively from (\ref
{ricci1}), (\ref{ricci2}), (\ref{ricci3}), (\ref{ricci4}).

We note that we can define the N--coefficients (\ref{ncoef}), $q_i(x^k,z)$
and $n_i(x^k,z),$ by solving the equations (\ref{einsteq3c}) and (\ref
{einsteq3d}) if the functions $h_i(x^k,z)$ are known as solutions of the
equations (\ref{einsteq3b}).

An elongated rotation ellipsoid hypersurface is given by the formula \cite
{korn}
\begin{equation}
\label{relhor}\frac{\widetilde{x}^2+\widetilde{y}^2}{\sigma ^2-1}+\frac{%
\widetilde{z}^2}{\sigma ^2}=\widetilde{\rho }^2,
\end{equation}
where $\sigma \geq 1$ and $\widetilde{\rho }$ is similar to the radial
coordinate in the spherical symmetric case.

The space 3D coordinate system is defined%
\begin{eqnarray}
\widetilde{x} &=&\widetilde{\rho}\sinh u\sin v\cos \varphi ,\
\widetilde{y}=\widetilde{\rho}\sinh u\sin v\sin \varphi ,\nonumber \\
\widetilde{z}&=& \widetilde{\rho}\ \cosh u\cos v, \nonumber
\end{eqnarray}
where $\sigma =\cosh u,(0\leq u<\infty ,\ 0\leq v\leq \pi ,\ 0\leq \varphi
<2\pi ). $\ The hypersurface metric is
\begin{eqnarray}
g_{uu} &=& g_{vv}=\widetilde{\rho}^2\left( \sinh ^2u+\sin ^2v\right) ,
 \label{hsuf1} \\
g_{\varphi \varphi } &=&\widetilde{\rho}^2\sinh ^2u\sin ^2v.
 \nonumber
\end{eqnarray}

Let us introduce a d--metric
\begin{equation}
\label{rel1}\delta s^2 = g_1(u,v)du^2+dv^2 + 
 h_3\left( u,v,\varphi \right) \left( \delta t\right) ^2+
h_4\left( u,v,\varphi \right) \left( \delta\varphi \right) ^2,
 \end{equation}
where $\delta t$ and $\delta \varphi $ are N--elongated differentials.

As a particular solution (\ref{einsteq3a}) for the h--metric, considering $%
\Upsilon _3^3=\Upsilon _4^4,$ we choose the coefficient
\begin{equation}
\label{relh1h}g_1(u,v)=\cos ^2v.
\end{equation}
The $h_3(u,v,\varphi )=h_3(u,v,\widetilde{\rho }\left( u,v,\varphi \right) )$
is considered as
\begin{equation}
\label{relh1}h_3(u,v,\widetilde{\rho })=\frac 1{\sinh ^2u+\sin ^2v}\frac{%
\left[ 1-\frac{r_g}{4\widetilde{\rho }}\right] ^2}{\left[ 1+\frac{r_g}{4%
\widetilde{\rho }}\right] ^6}.
\end{equation}
In order to define the $h_4$ coefficient solving the Einstein equations, for
simplicity, with a diagonal energy--momentum d--tensor for vanishing
pressure, we must solve the equation (\ref{einsteq3b}) which transforms into
a linear equation if $\Upsilon _1=0.$ In our case $s\left( u,v,\varphi
\right) =\beta ^{-1}\left( u,v,\varphi \right) ,$ where $\beta =\left(
\partial h_4/\partial \varphi \right) /h_4,$ must be a solution of
$$
\frac{\partial s}{\partial \varphi }+\frac{\partial \ln \sqrt{\left|
h_3\right| }}{\partial \varphi }\ s=\frac 12.
$$
After two integrations (see \cite{kamke}) the general solution for $%
h_4(u,v,\varphi ),$ is
\begin{equation}
\label{relh1a}h_4(u,v,\varphi )=a_4\left( u,v\right) \exp \left[
-\int\limits_0^\varphi F(u,v,z)\ dz\right] ,
\end{equation}
where%
\begin{equation}
F(u,v,z)= (\sqrt{|h_3(u,v,z)|}[s_{1(0)}\left( u,v\right)  +
\frac 12\int\limits_{z_0\left( u,v\right) }^z\sqrt{|h_3(u,v,z)|}dz])^{-1},
 \nonumber
\end{equation}
$s_{1(0)}\left( u,v\right) $ and $z_0\left( u,v\right) $ are some functions
of necessary smooth class. We note that if we put $h_4=a_4(u,v)$ the
equations (\ref{einsteq3b}) are satisfied for every $h_3=h_3(u,v,\varphi ).$

Every d--metric (\ref{rel1}) with coefficients of type (\ref{relh1h}), (\ref
{relh1}) and (\ref{relh1a}) solves the Einstein equations (\ref{einsteq3a}%
)--(\ref{einsteq3d}) with the diagonal momentum d--tensor
$$
\Upsilon _\beta ^\alpha =diag\left[ 0,0,-\varepsilon =-m_0,0\right] ,
$$
when $r_g=2\kappa m_0;$ we set the light constant $c=1.$ If we choose
$$
a_4\left( u,v\right) =\frac{\sinh ^2u\ \sin ^2v}{\sinh ^2u+\sin ^2v}
$$
our solution is conformally equivalent (if not considering the time--time
component) to the hypersurface metric (\ref{hsuf1}). The condition of
vanishing of the coefficient (\ref{relh1}) parametrizes the rotation
ellipsoid for the horizon
\begin{equation}
\label{ele4}\frac{\widetilde{x}^2+\widetilde{y}^2}{\sigma ^2-1}+\frac{%
\widetilde{z}^2}{\sigma ^2}=\left( \frac{r_g}4\right) ^2,
\end{equation}
where the radial coordinate is redefined via relation\ $\widetilde{r}=%
\widetilde{\rho }\left( 1+\frac{r_g}{4\widetilde{\rho }}\right) ^2. $ After
multiplication on the conformal factor
$$
\left( \sinh ^2u+\sin ^2v\right) \left[ 1+\frac{r_g}{4\widetilde{\rho }}%
\right] ^4,
$$
approximating $g_1(u,v)=\sin ^2v\approx 0,$ in the limit of locally
isotropic spherical symmetry,%
$$
\widetilde{x}^2+\widetilde{y}^2+\widetilde{z}^2=r_g^2,
$$
the d--metric (\ref{rel1}) reduces to
$$
ds^2=\left[ 1+\frac{r_g}{4\widetilde{\rho }}\right] ^4\left( d\widetilde{x}%
^2+d\widetilde{y}^2+d\widetilde{z}^2\right) -\frac{\left[ 1-\frac{r_g}{4%
\widetilde{\rho }}\right] ^2}{\left[ 1+\frac{r_g}{4\widetilde{\rho }}\right]
^2}dt^2
$$
which is just the Schwazschild solution with the redefined radial coordinate
when the space component becomes conformally Euclidean.

So, the d--metric (\ref{rel1}), the coefficients of N--connection being
solutions of (\ref{einsteq3c}) and (\ref{einsteq3d}), describe a static 4D
solution of the Einstein equations when instead of a spherical symmetric
horizon one considers a locally anisotropic deformation to the hypersurface
of rotation elongated ellipsoid.

\end{document}